\documentclass[twocolumn,prb,reprint,amsmath,amssymb,aps,showpacs]{revtex4-1}

\usepackage{graphicx}
\usepackage{dcolumn}
\usepackage{bm}
\usepackage{graphics}
\usepackage{epsfig}
\usepackage{epstopdf}
\begin{document}

\title{The influence of the Rashba spin-orbit coupling on the two-dimensional electron-hole system and magnetoexcitons}

\author{S.A. Moskalenko,$^{1}$ I.V. Podlesny,$^{1}$ E.V. Dumanov,$^{1,2}$
and A.A. Kiselyov$^{3}$}
\affiliation{$^{1}$Institute of Applied Physics of the Academy of Sciences of Moldova, Academic Str. 5, Chisinau, MD2028, Republic of Moldova\\
$^{2}$Technical University of Moldova, bd. Stefan cel Mare 168, MD2004, Chisinau, Republic of Moldova\\
$^{3}$State University of Civil Aviation, 38, Pilotov str., 196210, St. Petersburg, Russia}

\date{\today}

\begin{abstract}
In the review article the influence of the Rashba spin-orbit coupling on the two-dimensional electrons and holes in a strong perpendicular magnetic field is described in Landau gauge using the spinor-type two-component wave functions with different numbers of the Landau quantization levels in different spin projections, their difference being determined by the order of the chirality terms. In this representation the two-particle integral operators such as the electron and hole densities ${{\hat{\rho }}_{e}}(\vec{Q})$, and ${{\hat{\rho }}_{h}}(\vec{Q})$ as well as the magnetoexciton creation and annihilation operators $\hat{\Psi }_{ex}^{\dagger }({{F}_{n}},\vec{k})$ and $\hat{\Psi }_{ex}^{{}}({{F}_{n}},\vec{k})$ in different combinations ${{F}_{n}}$ of the electron and hole spinor states were introduced. On this base the Hamiltonians of the Coulomb electron-electron and electron-radiation interactions were deduced. The properties of the magnetoexcitons in these conditions were discussed.
\end{abstract}

\maketitle
\tableofcontents


\section{Introduction}

The present review article is based on the background previous papers and monographs [1-16] as well on the recent contribution [17-25].
	In Refs.[6, 19] it was underlined that the spin degeneracy of the electron and hole states is the combined effect of the inversion symmetry in space and time. The first operator is denoted by $\hat{I}$ and the second one by $K=\sigma_2K_0$, where $\sigma_2$ is the Pauli matrix and $K_0$ means the complex conjugation operation. They change the Bloch wave functions in solids characterized by wave vector $\vec{k}$ and spin index $\sigma$ with two projections in the following way $I{{\Psi }_{\sigma }}(\vec{r},\vec{k})={{\Psi }_{\sigma }}(-\vec{r},\vec{k})={{\Psi }_{\sigma }}(\vec{r},-\vec{k})$, $\sigma =\uparrow ,\downarrow $, $\hat{K}{{\Psi }_{\uparrow }}(\vec{r},\vec{k})=\Psi _{\downarrow }^{*}(\vec{r},\vec{k})={{\Psi }_{\downarrow }}(\vec{r},-\vec{k})$, ${{\hat{K}}^{2}}=-\hat{1}$.
The time inversion operator $\hat{K}$ flips the spin side by side with the complex conjugation operation.
The first symmetry operator means the equality$E_{\sigma} (\vec{k}) = E_{\sigma} (-\vec{k}),$whereas the time inversion operator leads to Kramers degeneracy ${{E}_{\uparrow }}(\vec{k})={{E}_{\downarrow }}(-\vec{k})$.
It takes place, even if the space inversion is absent. The invariance of the Hamiltonian under the action of two inversion operations leads to two-fold spin degeneracy of the single-particle states with arbitrary wave vector $\vec{k}$ as follows ${{E}_{\uparrow }}(\vec{k})={{E}_{\uparrow }}(-\vec{k})={{E}_{\downarrow }}(-\vec{k})$, ${{E}_{\uparrow }}(\vec{k})={{E}_{\downarrow }}(\vec{k})$.
	These relations are true for both electrons and holes. Nevertheless the Rashba spin splitting of 2D hole systems is very different from the more familiar case of 2D electron systems. In [6, 19] it was explained by the fact that the holes have typically larger masses and smaller kinetic energies. The SOC is more important for holes than for electrons. When the carriers are moving through the inversion asymmetric potential, the spin degeneracy is removed even in the absence of an external magnetic field $H$. In this case there are two different branches of energy $E_{\uparrow} (\vec{k}) \ne E_{\downarrow} (\vec{k})$ and the spin splitting is present. In quasi-2D QWs this spin splitting can be the consequence of a bulk inversion asymmetry(BIA) of the underlying crystal (for example, as in zinc blende crystal), or of a structure inversion asymmetry(SIA) of the confinement potential.
	In both cases of inversion asymmetry the spin splitting takes place in the absence of $H$, i.e. $E_{\uparrow} (\vec{k}) \ne E_{\downarrow} (\vec{k})$, but the Kramers degeneracy continues to exist $E_{\uparrow} (\vec{k}) = E_{\downarrow} (-\vec{k})$. This spin splitting is not due to Zeeman effect because $H=0$. In [6, 19] the origin of the spin splitting is related with the motion of the electron through the inversion asymmetric spatial environment, the interaction with which is due to the SOC. The periodic part of the electron Bloch functions are feeling the atomic fields, that enter into the Pauli SO term, whereas the envelope functions feel the macroscopic environment. Following this picture, SIA leads to spin splitting, which is due to both macroscopic electric field and microscopic electric field from the atomic cores.
	SIA spin splitting is always proportional to the macroscopic field strength times a prefactor depending on the microscopic spin-orbit interaction (SOI). This prefactor depends only on the matrix elements of the microscopic SOI and is due completely to the BIA. To reveal the origin of the spin splitting in a more simple way the following idea was suggested. One can imagine the electron moving with velocity $V_{||}$ in the plane of the layer subjected to the action of a perpendicular electric field $E_z$.
	In the reference frame moving together with the electron the Lorentz transformation induces the magnetic field $H=({{V}_{||}}/c){{E}_{z}}$, which acts on the electron spin giving rise to such indirect Zeeman effect. The made estimations showed that the obtained in such a way spin splitting is by 5-6 orders of magnitude smaller than the experimentally observed values of the SOC. The discrepancy is due to the fact that the idea of Lorentz transformation neglects the contribution of the atomic cores to the SOI felt by Bloch electrons in a solid [6, 19]. Another important detail, which must be remembered is related with the crystallographic symmetry group of the solids. The spin splitting induced by the atomic cores, which is named also as BIA splitting, does depend also on the irreducible representations of the double group of the wave vector $\vec{k}$. For example, in the case of $\vec{k}$ parallel to $\left\langle 111 \right\rangle $ direction the wave vector group is $\mathrm{C_{3v}}$. It has a double-group irreducible representations $\mathrm{\Gamma_{4}}$, $\mathrm{\Gamma_{5}}$ and $\mathrm{\Gamma_{6}}$. If the electron and light-hole(LH) states transform according the two-dimensional representations $\mathrm{\Gamma_{4}}$, whereas the heavy-hole(HH) states transform according the one-dimensional representations $\mathrm{\Gamma_{5}}$ and $\mathrm{\Gamma_{6}}$, in this case the BIA spin splitting vanishes for electrons and LHs and does exist for HHs [6, 19]. RSOC and intrinsic SOI under certain conditions lead to a Dirac cone formation out of a parabolic band and it is possible to create a "Mexican-hatlike" energy dispersion law [8, 19]. The Mexican-hatlike dispersion has a line of degenerate low-energy points forming a ring. It can appear in a variety of physical systems. Such peculiarities were demonstrated in [2, 3, 19]. The Mexican-hatlike dispersion law leads to a weak crystallization transition, [9, 19] whereas in the cold atoms physics it gives rise to topologically different ground states of the Bose-Einstein condensed atoms and molecules [8, 19].
	In Ref.[17] the Hamiltonian of the electron-radiation interaction in the second quantization representation for the case of 2D coplanar electron-hole (e-h) systems in a strong perpendicular magnetic field was derived. The s-type conduction band electrons with spin projections ${{s}_{z}}=\pm 1/2$ along the magnetic field direction and the heavy holes with total momentum projections ${{j}_{z}}=\pm 3/2$ in the p-type valence band were taken into account. The periodic parts of their Bloch wave functions are similar to $(x\pm iy)$ expressions with the orbital momentum projection ${{M}_{v}}=\pm 1$ on the same selected direction. The envelope parts of the Bloch wave functions have the forms of plane waves in the absence of the magnetic field. In its presence they a completely changed due to the Landau quantization event. In the papers [17-25] the Landau quantization of the 2D electrons and holes is described in the Landau gauge and is characterized by the oscillator-type motion in one in-plane direction giving rise to discrete Landau levels enumerated by the quantum numbers ${{n}_{e}}$ and ${{n}_{h}}$ and by the free translation motion in another in-plane direction perpendicular to previous one. The one-dimensional (1D) plane waves describing this motion are marked by the 1D wave numbers p and q. In Ref.[18] the Landau quantization of the 2D electrons with non-parabolic dispersion law, pseudospin components and chirality terms was investigated. On this base in Ref.[19] the influence of the Rashba spin-orbit coupling (RSOC) on the 2D magnetoexcitons was discussed. The spinor-type wave functions of the conduction and valence electrons in the presence of the RSOC have different numbers of Landau quantization levels for different spin projections. As was  demonstrated in Refs[18, 19, 22], the difference between these numbers is determined by the order of the chirality terms. Their origin is due to the influence of the external electric field applied to the layer parallel to the direction of the magnetic field. In Ref.[19] two lowest Landau levels (LLLs) of the conduction electron and four LLLs for the holes were used to calculate the matrix elements of the Coulomb interaction between the charged carriers as well as the matrix elements of the electron-radiation interaction. On these bases the ionization potentials of the new-magnetoexcitons and the probabilities of the quantum transitions from the ground state of the crystal to the magnetoexciton states were calculated. In the present description the number of the hole and magnetoexciton states will be enlarged and the formation of magnetopolaritons taking into account the RSOC will be described. The more simple variant of the magnetopolariton without taking into account the RSOC was described in Ref.[21] for the case of inter-band quantum transitions and in Ref.[23] for the case of intra-band quantum transitions.
	First of all we will describe the Landau quantization of the 2D heavy holes following the Ref.[19, 22].
	The review article is organized as follows. In the section 2 the Landau quantization of the heavy holes taking into account the RSOC is described in details. On this base as well as using the Rashba results for the conduction electrons [1] the Hamiltonian of the electron-radiation interaction was deduced in Section 3. The Hamiltonian of the Coulomb electron-electron interaction is presented in Section 4. The Section 5 is devoted to the conclusions.

\section{Landau quantization of the 2D heavy holes}
The full Landau-Rashba Hamiltonian for 2D heavy holes was discussed in Ref.[19] following the formulas (13)-(20). It can be expressed through the Bose-type creation and annihilation operators ${{a}^{\dagger }}$, $a$ acting on the Fock quantum states $\left| n \right\rangle =\frac{{{({{a}^{\dagger }})}^{n}}}{\sqrt{n!}}\left| 0 \right\rangle $, where $\left| 0 \right \rangle$ is the vacuum state of harmonic oscillator. The Hamiltonian has the form [22]
\begin{eqnarray}
\hat{H}_h &=& \hbar \omega_{ch} \left\lbrace \left[ \left( a^{\dag}a + \frac{1}{2}\right) + \delta \left( a^{\dag}a + \frac{1}{2}\right)^2 \right] \hat{I} \right. \nonumber \\
&& \left. + i \beta 2\sqrt{2} \left|\begin{array}{cc}
0 & (a^{\dag})^3\\
-a^3 & 0\end{array}\right| \right\rbrace; \hat{I} = \left|\begin{array}{cc}
1 & 0\\
0 & 1\end{array}\right| \label{eq:fullhameh}
\end{eqnarray}
with the denotations

\begin{eqnarray}
& \omega_{ch} = \frac{|e|H}{m_h c}; \: \delta = \frac{\left| \delta_h E_z\right| \hbar^4}{l^4 \hbar\omega_{ch}}; \: \beta = \frac{\beta_{h}E_{z}}{l^{3}\hbar\omega_{ch}}; \: l = \sqrt{\frac{\hbar c}{|e|H}}&. \label{eq:denotwdbl}
\end{eqnarray}
The parameter $\delta_h$ is not well known, what permits to consider different variants mentioned below.

The exact solutions of the Pauli-type Hamiltonian is described by the formulas (21)-(31) of the Ref.[19]. In more details they were described in Ref.[22], and have the spinor form
\begin{eqnarray}
& \hat{H}_h \left|\begin{array}{c}
f_{1}\\
f_{2}\end{array}\right| = E_{h} \left|\begin{array}{c}
f_{1}\\
f_{2}\end{array}\right|; \: f_1 = {\displaystyle \sum_{n=0}^{\infty} c_n \left| n \right \rangle}; \; f_2 = {\displaystyle \sum_{n=0}^{\infty} d_n \left| n \right \rangle}; & \nonumber \\
& {\displaystyle \sum_{n=0}^{\infty} |c_n|^2 + \sum_{n=0}^{\infty} |d_n|^2 = 1}. & \label{eq:solptham}
\end{eqnarray}
First three solutions depend only on one quantum number $m$ with the values $0, 1, 2$ as follows

\begin{eqnarray}
& E_h(m=0) = \hbar \omega_{ch} \left( \frac12 + \delta \right); \: \Psi (m=0) = \left|\begin{array}{c}
\left| 0 \right \rangle \\
0 \end{array} \right|, & \nonumber \\
& E_h(m=1) = \hbar \omega_{ch} \left( \frac32 + 9 \delta \right); \: \Psi (m=1) = \left|\begin{array}{c}
\left| 1 \right \rangle \\
0 \end{array} \right|, & \nonumber \\
& E_h(m=2) = \hbar \omega_{ch} \left( \frac52 + 25 \delta \right); \: \Psi (m=2) = \left|\begin{array}{c}
\left| 2 \right \rangle \\
0 \end{array}\right|. & \label{eq:enhthreestates}
\end{eqnarray}
All another solutions with $m \ge 3$ depend on two quantum numbers $(m-5/2)$ and $(m+1/2)$ and have the general expression

\begin{eqnarray}
& \varepsilon_{h}^{\pm} (m - \frac52; m + \frac12) = \frac{E_{h}^{\pm} (m-5/2; m+1/2)}{\hbar \omega_{ch}} & \nonumber \\
& = (m-1) + \frac{\delta}{2} \left[ (2m+1)^2 + (2m-5)^2\right] & \nonumber \\
& \pm (\left( \frac32 + \frac{\delta}{2} \left[ (2m+1)^2 - (2m-5)^2\right] \right)^2 & \nonumber \\
& + 8 \beta^2 m(m-1)(m-2))^{1/2}, \: m \ge 3. &
\label{eq:exactsoldl}
\end{eqnarray}
The corresponding wave functions for $m=3$ and $m=4$ are

\begin{eqnarray}
& \Psi_{h}^{\pm} (m=3) = \left|\begin{array}{c}
c_{3} \left| 3 \right\rangle \\
d_{0} \left| 0 \right\rangle \end{array}\right| \: \textrm{and} \:
\Psi_{h}^{\pm} (m=4) = \left|\begin{array}{c}
c_{4} \left| 4 \right\rangle \\
d_{1} \left| 1 \right\rangle \end{array}\right|. & \label{eq:hwfspipar}
\end{eqnarray}
They depend on the coefficients $c_m$ and $d_{m-3}$, which obey to the equations

\begin{eqnarray}
& c_m \left( m + \frac12 + \delta (2m+1)^2 - \varepsilon_{h} \right) & \nonumber \\
& = -i \beta 2 \sqrt{2} \sqrt{m(m-1)(m-2)} d_{m-3}; & \nonumber \\
& d_{m-3} \left( m - \frac52 + \delta (2m-5)^2 - \varepsilon_{h} \right) & \nonumber \\
& = i \beta 2 \sqrt{2} \sqrt{m(m-1)(m-2)} c_{m}; & \nonumber \\
& |c_m|^2 + |d_{m-3}|^2 = 1.& \label{eq:displawhsyst}
\end{eqnarray}
There are two different solutions $\varepsilon_{h}^{\pm} (m)$ at a given value of $m \ge 3$ and two different pairs of the coefficients $(c_m^{\pm}, d_{m-3}^{\pm})$.
The dependences of the parameters $\omega_{ch}$, $\beta$ and $\delta$ on the electric and magnetic fields strengths may be represented for the GaAs-type quantum wells as follows $H = y \; \mathrm{T}$; $E_z = x \; \mathrm{kV/cm}$; $m_h = 0.25 m_0$; $\hbar \omega_{ch} = 0.4 y \; \mathrm{meV}$; $\beta = 1.062 \cdot 10^{-2} x \sqrt{y}$; $\delta = 10^{-4} C x y$ with unknown parameter $C$, which will be varied in more large interval of values. We cannot neglect the parameter $C$ putting it equal to zero, because in this case, as was argued in Ref.[19] formula (10), the lower spinor branch of the heavy hole dispersion law
\begin{eqnarray*}
& E_{h}^{-} (k_{||}) = \frac{\hbar^2 \vec{k}_{||}^2}{2m_h} - \left| \frac{\beta_h E_z}{2} \right| \left| \vec{k}_{||} \right|^3 & \label{eq:tspinbr}
\end{eqnarray*}
has an unlimited decreasing, deeply penetrating inside the semiconductor energy gap at great values of $\left| \vec{k}_{||} \right|$. To avoid this unphysical situation the positive quartic term $\left| \delta_hE_z \right| \vec{k}_{||}^4$ was added in the starting Hamiltonian. The new dependences were compared with the drawings calculated in the fig.2 of the Ref.[19] in the case $E_z = 10 \; \mathrm{kV/cm}$ and $C=10$. Four lowest Landau levels(LLLs) for heavy holes were selected in Ref.[19]. In addition to them in Ref.[22] were studied else three levels as follows
\begin{eqnarray}
& E_h(R_1) = E_h^{-} (\frac12, \frac72); \: E_h(R_2) = E_h(m=0); & \nonumber \\
& E_h(R_3) = E_h^{-} (\frac32, \frac92); \: E_h(R_4) = E_h(m=1); & \nonumber \\
& E_h(R_5) = E_h^{-} (\frac52, \frac{11}{2}); \: E_h(R_6) = E_h(m=2); & \nonumber \\
& E_h(R_7) = E_h^{-} (\frac72, \frac{13}{2}). &
\end{eqnarray}
Their dependences on the magnetic field strength were represented in the figures 1 and 2 of the Ref.[22] at different parameters $x$ and $C$ and are reproduced below.
\begin{figure}[h]
\includegraphics[scale=0.55]{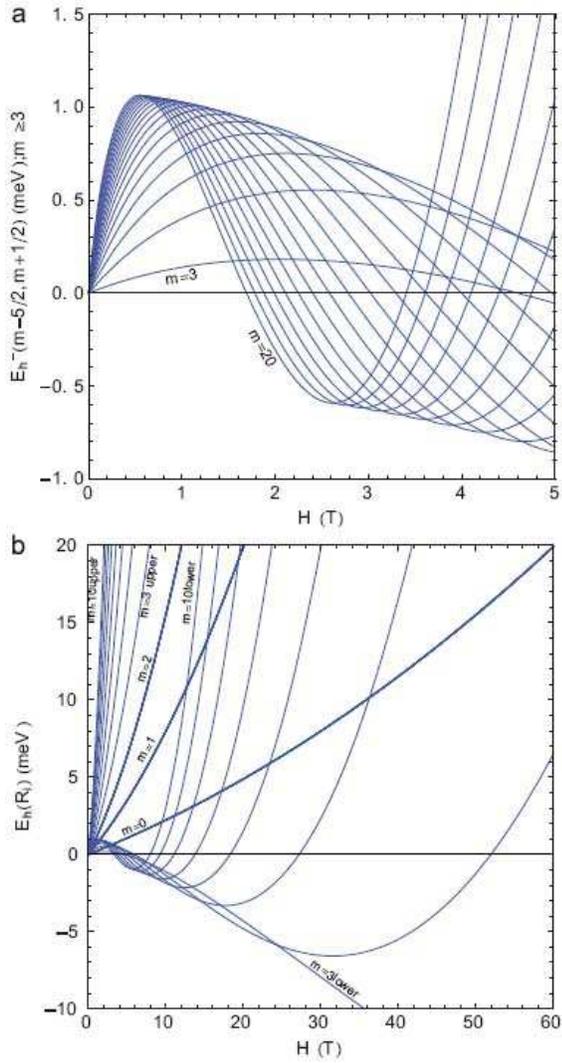}
\caption{a) The lower branches of the heavy hole Landau quantization levels $E_{h}^{-} (m-5/2; m+1/2)$ for $m \ge 3$ at the parameters $E_z = 10 \; \mathrm{kV/cm}$ and $C=5.5$; b) The general view of the all heavy hole Landau quantization levels with m=0,1,...,10 at the same parameters $E_z$ and $C$. They are reproduced from the fig.1 of the Ref.[22].}
\end{figure}

\begin{figure}[h]
\includegraphics[scale=0.36]{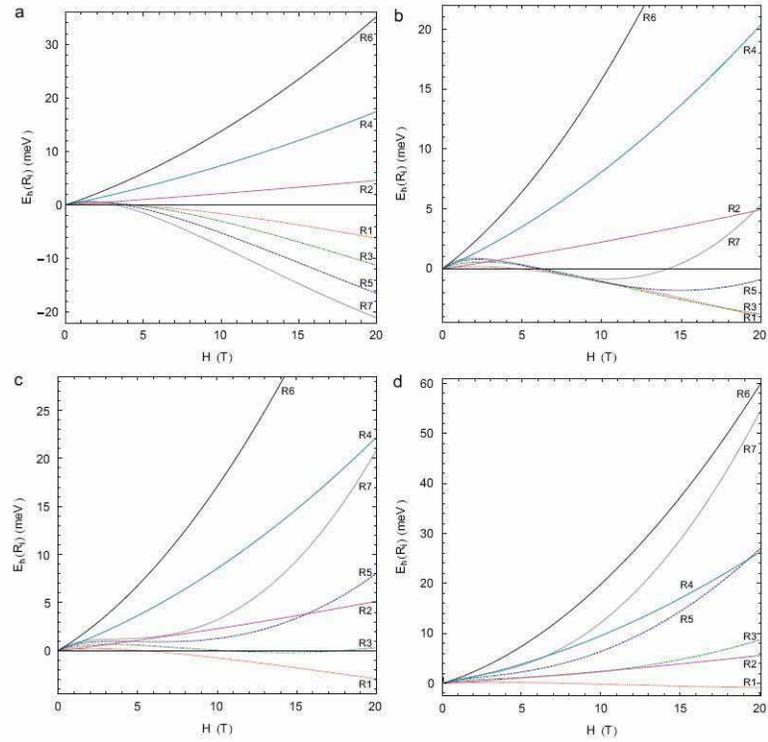}
\caption{Seven branches of the heavy hole Landau quantization levels with $m=0,1,\ldots,6$ at the parameter $E_z = 10 \; \mathrm{kV/cm}$ and different values of the parameter $C$: 3.8(a), 5.8(b), 7.05(c) and 10(d). They are reproduced from fig.2 of the Ref.[22].}
\end{figure}

The general view of the lower branches $E_h^{-}(m-\frac52, m+\frac12)$ of the heavy hole Landau quantization levels with $m \ge 3$ as a functions of the magnetic field strength are represented in fig.1a following the formula (5). The upper branches have more simple monotonous behavior and are drawn in fig.1b together with some curves of the lower branches. All the lower branches in their initial parts have a linear increasing behavior up till they achieve the maximal values succeeded by the minimal values in the middle parts of their evolutions being transformed in the final quadratic increasing dependences. The values of the magnetic field strength corresponding to the minima and to the maxima decrease with the increasing of the number $m$. These peculiarities can be compared with the case of Landau quantization of the 2D electron in the biased bilayer graphene described in Ref.[18]. The last case is characterized by the initial dispersion law without parabolic part and by second order chirality terms. They both lead to dependences on the magnetic field strength for the lower dispersion branches with sharp initial decreasing parts and minimal values succeeded by the quadratic increasing behavior. The differences between the initial dispersion laws and chirality terms in two cases of bilayer graphene and of heavy holes lead to different intersections and degeneracies of the Landau levels. The fig.2 shows that the change of the parameter $C$ at a given parameter $E_z$ (or vice versa) shifts significantly on the energy scale the lower branches of the heavy hole Landau levels. It can be observed in all four sections of the fig.2. But there is a special case in the section 2b, where the degeneracy of the levels $R_1$, $R_3$, $R_5$ and $R_7$ take place in the range of magnetic field strength $(5-10) \; \mathrm{T}$. More so, the degeneracy of the levels $R_1$ and $R_3$ persists to exist even in a more large interval as $(5-20) \; \mathrm{T}$.
	The degeneracy of two lowest Landau levels(LLLs) in biased bilayer graphene was suggested to explain the experimental results related with the fractional quantum Hall effects [15, 18]. Meanwhile the degeneracy of two LLLs in biased bilayer graphene in Ref.[18] was revealed only near the intersection point at a given value of the magnetic field strength. To obtain a more complete and wide degeneracy in the calculations concerning the biased bilayer graphene one could employ a mixed model using the results reflected in fig.2.
	The spinor-type envelope wave fynctions of the heavy holes in the coordinate representation look as [22]
\begin{eqnarray}
  |\psi_{h}(\varepsilon_{m},q;x,y)\rangle = \frac{e^{iqx}}{\sqrt{L_{x}}}\left\vert
\begin{array}{c}
\varphi_{m}(y+ql^{2}) \\
0%
\end{array}%
\right\vert ,\nonumber \\
  m=0,1,2, \\
|\psi_{h}(\varepsilon^{\pm}_{m},q;x,y)\rangle =  \frac{e^{iqx}}{\sqrt{L_{x}}}\left\vert
\begin{array}{c}
c^{\pm}_{m}\varphi_{m}(y+ql^{2}) \\
d^{\pm}_{m-3}\varphi_{m-3}(y+ql^{2})%
\end{array}%
\right\vert , \nonumber \\
  m \geq 3. \nonumber
\end{eqnarray}
The valence electrons in comparison with the holes are characterized by the opposite signs of he spin projections, wave vector and charge. The corresponding envelope wave functions can be obtained from the previous ones by the procedure
\begin{equation}
\left\vert \psi _{h}(\varepsilon _{m},q;x,y)\right\rangle =i\widehat{\sigma }_{y}\left\vert \psi _{h}(\varepsilon _{m},-q;x,y)\right\rangle ^{\ast }
\end{equation}
where ${{\hat{\sigma }}_{y}}$ is the Pauli matrix. In coordinate representation they are
\begin{eqnarray}
  |\psi_{v}(\varepsilon_{m},q;x,y)\rangle = \frac{e^{iqx}}{\sqrt{L_{x}}}\left\vert
\begin{array}{c}
0 \\
-\varphi^{*}_{m}(y-ql^{2})%
\end{array}%
\right\vert ,\nonumber \\
  m=0,1,2, \\
|\psi_{h}(\varepsilon^{\pm}_{m},q;x,y)\rangle =  \frac{e^{iqx}}{\sqrt{L_{x}}}\left\vert
\begin{array}{c}
d^{\pm *}_{m-3}\varphi^{*}_{m-3}(y-ql^{2}) \\
-c^{\pm *}_{m}\varphi^{*}_{m}(y-ql^{2})%
\end{array}%
\right\vert , \nonumber \\
  m \geq 3. \nonumber
\end{eqnarray}
To obtain the full valence electron Bloch wave functions the expressions (11) must be multiplied by the periodic parts. In the p-type valence band they have the form $\frac{1}{\sqrt{2}}({{U}_{v,p,x,q}}(x,y)\pm i{{U}_{v,p,y,q}}(x,y))$ and are characterized by the orbital momentum projections ${{M}_{v}}=\pm 1$ correspondingly. The hole orbital projections ${{M}_{h}}=-{{M}_{v}}$ have opposite signs in comparison with the valence electron. The full Bloch wave functions of the valence electrons now are characterized by a supplementary quantum number ${{M}_{v}}$ side by side with the previous ones ${{\varepsilon }_{m}}$, ${{\varepsilon }^{\pm }}_{m}$ and q as follows
\begin{eqnarray}
\left\vert \psi _{v}\left( M_{v},\varepsilon _{m},q;x,y\right) \right\rangle
&=&\frac{e^{iqx}}{\sqrt{L_{x}}}\frac{1}{\sqrt{2}}(U_{v,p,x,q}(\vec{r})
\nonumber \\
&&\pm iU_{v,p,y,q}(\vec{r}))\left\vert
\begin{array}{c}
0 \\
-\varphi _{m}^{\ast }(y-ql^{2})%
\end{array}%
\right\vert ;  \nonumber \\
m &=&0,1,2;M_{v}=\pm 1, \\
\left\vert \psi _{v}\left( M_{v},\varepsilon _{m}^{\pm },q;x,y\right)
\right\rangle  &=&\frac{e^{iqx}}{\sqrt{L_{x}}}\frac{1}{\sqrt{2}}(U_{v,p,x,q}(%
\vec{r})  \nonumber \\
&&\pm iU_{v,p,y,q}(\vec{r}))\left\vert
\begin{array}{c}
d_{m-3}^{\pm \ast }\varphi _{m-3}^{\ast }(y-ql^{2}) \\
-c_{m}^{\pm \ast }\varphi _{m}^{\ast }(y-ql^{2})%
\end{array}%
\right\vert ,  \nonumber \\
m &\geq &3,M_{v}=\pm 1  \nonumber
\end{eqnarray}
From this multitude of valence electron wave functions the more important of them are characterized by the values $\varepsilon _{m}^{-}$ with $m=3$ and 4, aw well as by $\varepsilon _{m}^{{}}$ with $m=0,1$. These four lowest hole energy levels being combined with two projections ${{M}_{h}}\pm 1$ form a set of 8 lowest hole states, which will be taken into account below.

Now for the completeness we will remember the main results obtained by Rashba [1] in the case of the electron conduction band. They are needed to obtain a full description of the 2D electron-hole pair and of a 2D magnetoexciton in the condition of the Landau quantization under the influence of the RSOC.

The lowest Landau level of the conduction electron in the presence of the RSOC was obtained in Ref.[1]:
\begin{eqnarray}
& \left| \psi_{\mathrm{e}}\left( R_{1},p;x_e,y_e \right)\right\rangle = \frac{e^{ipx_{e}}}{\sqrt{L_{x}}} \left|\begin{array}{c}
a_{0}\varphi_{0}(y_{\mathrm{e}})\\
b_{1}\varphi_{1}(y_{\mathrm{e}})\end{array}\right|; & \nonumber \\
& \varepsilon_{{\mathrm{e}}R_{1}} = 1 - \sqrt{\frac{1}{4} + 2\alpha^{2}}; \: |a_{0}|^{2} + |b_{1}|^{2} = 1 & \nonumber \\
& |a_{0}|^{2} = \frac{1}{1 + \frac{2\alpha^{2}}{\left[\frac{1}{2} + \sqrt{\frac{1}{4} + 2\alpha^{2}}\right]^{2}}}; \: |b_{1}|^{2} = \frac{2\alpha^{2}|a_{0}|^{2}}{\left[\frac{1}{2} + \sqrt{\frac{1}{4} + 2\alpha^{2}}\right]^{2}}.
\end{eqnarray}
The next electron level higher situated on the energy scale is characterized by the pure spin oriented state
\begin{eqnarray}
& \left| \psi_{\mathrm{e}}\left( R_{2},p;x_{\mathrm{e}},y_{\mathrm{e}} \right)\right\rangle = \frac{e^{ipx_{e}}}{\sqrt{L_{x}}} \left|\begin{array}{c}
0\\
\varphi_{0}(y_{\mathrm{e}})\end{array}\right|; \: \varepsilon_{{\mathrm{e}}R_{2}} = \frac12. &
\end{eqnarray}
Two lowest Landau levels~(LLLs) for conduction electron are characterized by the values $m_{\mathrm{e}} = 0.067 m_0$, $\hbar \omega_{\mathrm{ce}} = 1.49 \; \mathrm{meV}  \cdot y$ and parameter $\alpha = 8 \cdot 10^{-3} x/ \sqrt{y}$. They are denoted as

\begin{eqnarray}
& E_{\mathrm{e}}(R_{1}) = \hbar \omega_{\mathrm{ce}} \left( 1 - \sqrt{\frac{1}{4} + 2\alpha^{2}} \right); \nonumber \\
& E_{\mathrm{e}}(R_{2}) = \hbar \omega_{\mathrm{ce}} \frac12. &
\end{eqnarray}

The lowest Landau energy level for electron $E_{\mathrm{e}}(R_1)$ has a nonmonotonous anomalous dependence on the magnetic field strength near the point $H=0 \: \mathrm{T}$. It is due to the singular dependence of the RSOC parameter $\alpha^2 = 6.4 \cdot 10^{-5} x^2/y$, which is compensated in the total energy level expression by the factor $\hbar \omega_{\mathrm{ce}}$ of the cyclotron energy, where $\hbar \omega_{\mathrm{ce}} = 1.49 \: y$~meV. The second electron Landau energy level has a simple linear dependence on $H$.

The full Bloch wave functions for conduction electrons including their s-type periodic parts look as
\begin{eqnarray}
\left\vert \psi _{c}\left( s,R_{1},p;x,y\right) \right\rangle  &=&\frac{%
e^{ipx}}{\sqrt{L_{x}}}U_{c,s,p}(\vec{r})\left\vert
\begin{array}{c}
a_{0}\varphi _{0}(y-pl^{2}) \\
b_{1}\varphi _{1}(y-pl^{2})%
\end{array}%
\right\vert ,  \nonumber \\
\left\vert \psi _{c}\left( s,R_{2},p;x,y\right) \right\rangle  &=&\frac{%
e^{ipx}}{\sqrt{L_{x}}}U_{c,s,p}(\vec{r})\left\vert
\begin{array}{c}
0 \\
\varphi _{m}(y-pl^{2})%
\end{array}%
\right\vert .
\end{eqnarray}
Two lowest Rashba-type states for conduction electron will be combined with eight LLLs for heavy holes and with the corresponding states of the valence electrons. The e-h pair will be characterized by 16 states. Heaving the full set of the electron Bloch wave functions in conduction and in the valence bands one can construct the Hamiltonian describing in second quantization representation the Coulomb electron-electron interaction as well as the electron-radiation interaction. These tow tasks will be described in the next sections of our review paper. The results obtained earlier in the Ref.[19, 22] taking into account only 8 e-h states will be supplemented below.

\section{Electron-radiation interaction in the presence of the Rashba spin-orbit coupling}
In the Ref.[17, 21] the Hamiltonian of the electron-radiation interaction in the second quantization representation for the case of two-dimensional(2D) coplanar electron-hole(e-h) system in a strong perpendicular magnetic field was discussed. The $s$-type conduction-band electrons with spin projections $s_z = \pm 1/2$ along the magnetic field direction and the heavy holes with the total momentum projections $j_z = \pm 3/2$ in the $p$-type valence band were taken into account. Their orbital Bloch wave functions are similar to $(x \pm iy)$ expressions with the orbital momentum projections $M= \pm 1$ on the same selected direction. The Landau quantization of the 2D electrons and holes was described in the Landau gauge with oscillator type motion in one in-plane direction characterized by the quantum numbers $n_e$ and $n_h$ and with the free translational motion described by the uni-dimensional(1D) wave numbers $p$ and $q$ in another in-plane direction perpendicular to the previous one. The electron and hole creation and annihilation operators $a^{+}_{s_z, n_e, p}$, $a_{s_z, n_e, p}$, and $b^{+}_{j_z, n_h, q}$, $b_{j_z, n_h, q}$ were introduced correspondingly. The Zeeman effect and the Rashba spin-orbit coupling in Refs [17, 21] were not taken into account.
	The electrons and holes have a free orbital motion on the surface of the layer with the area $S$ and are completely confined in $\vec{a}_3$ direction. The degeneracy of their Landau levels equals to $N = S/(2 \pi l_0^2)$, where $l_0$ is the magnetic length. In contrast, the photons were supposed to move in any direction in the three-dimensional(3D) space with the wave vector $\vec{k}$ arbitrary oriented as regards the 2D layer as it is represented in the Fig.3 reproduced from the Ref.[17]. There are three unit vectors $\vec{a}_1$, $\vec{a}_2$, $\vec{a}_3$, the first two being in-plane oriented whereas the third $\vec{a}_3$ is perpendicular to the layer. We will use the 3D and 2D wave vectors $\vec{k}$ and $\vec{k}_{||}$ and will introduce the circular polarization vectors $\vec{\sigma}_{M}$ for the valence electrons, heavy holes and magnetoexcitons as follows
\begin{figure}[h]
\includegraphics[scale=0.35]{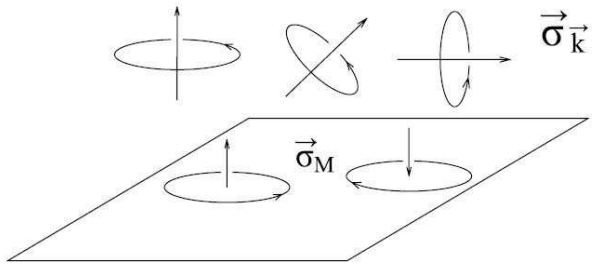}
\caption{The reciprocal orientations of the circularly polarized vectors $\vec{\sigma}_{\vec{k}}$ and $\vec{\sigma}_{M}$, reproduced from the Ref.[17].}
\end{figure}
\begin{eqnarray}
  \vec{k}=\vec{k}_{||}+\vec{a}_{3}k_{z}, \nonumber \\
  \vec{k}_{||}=\vec{a}_{1}k_{x}+\vec{a}_{2}k_{y}, \nonumber \\
  \vec{\sigma}_{M}=\frac{1}{\sqrt{2}}(\vec{a}_{1}\pm i\vec{a}_{2}), M=\pm 1.
\end{eqnarray}
The photons are characterized by two linear vectors $\vec{e}_{k,j}$ or by two circular polarization vectors $\vec{\sigma}_{\vec{k}}^{\pm}$ obeying the transversality conditions:
\begin{eqnarray}
\vec{\sigma }_{{\vec{k}}}^{\pm }=\frac{1}{\sqrt{2}}({{{\vec{e}}}_{\vec{k},1}}\pm i{{{\vec{e}}}_{\vec{k},2}}), \nonumber \\
 ({{{\vec{e}}}_{k,j}}\cdot \vec{k})=0; j=1,2.
\end{eqnarray}
The photon creation and annihilation operators can be introduced in two different polarizations as follows
\begin{eqnarray}
{{C}_{\vec{k},\pm }}=\frac{1}{\sqrt{2}}\left( {{C}_{\vec{k},1}}\pm i{{C}_{\vec{k},2}} \right), \nonumber \\
 {{\left( {{C}_{\vec{k},\pm }} \right)}^{\dagger }}=\frac{1}{\sqrt{2}}\left( C_{\vec{k},1}^{\dagger }\mp i C_{\vec{k},2}^{\dagger } \right), \nonumber \\
\sum^{2}_{j=1}{{\vec{e}}_{\vec{k},j}}{{C}_{\vec{k},j}}={{C}_{\vec{k},-}}\vec{\sigma }_{{\vec{k}}}^{+}+{{C}_{\vec{k},+}}\vec{\sigma }_{{\vec{k}}}^{-}, \nonumber \\
\sum^{2}_{j=1}{{\vec{e}}_{\vec{k},j}}C_{\vec{k},j}^{\dagger }={{\left( {{C}_{\vec{k},-}} \right)}^{\dagger }}\vec{\sigma }_{{\vec{k}}}^{-}+{{\left( {{C}_{\vec{k},+}} \right)}^{\dagger }}\vec{\sigma }_{{\vec{k}}}^{+}.
\end{eqnarray}
The reciprocal orientations of the circular polarizations $\vec{\sigma}_{\vec{k}}^{\pm}$ and $\vec{\sigma}_{M}$ will determine the values of the scalar products $(\vec{\sigma}_{\vec{k}}^{\pm} \cdot \vec{\sigma}_{M}^{\ast})$. The electron-radiation interaction describing only the band-to-band quantum transitions with the participation of the e-h pairs in the presence of a strong perpendicular magnetic field was obtained in Ref.[17] and can be used as initial expression for obtaining the interaction of 2D magnetoexcitons with the electromagnetic field. Following the formula (12) of Ref.[17] we have
\begin{widetext}
\begin{equation} \label{eq:hamabone}
\begin{array}{c}
  \displaystyle \hat{H}_{e-rad}=\left(-\frac{e}{m_0}\right)\sum_{\vec{k}(k_x, k_y, k_z)} \sqrt{\frac{2\pi\hbar}{V\omega_{k}}} \sum_{n_e, n_h} \sum_{p} \\
  \Bigg\{ P_{cv}(k_y, p)\Phi\left( n_e, p; n_h, p-k_x; k_y\right)
  \Bigg( \bigg[ C_{\vec{k}, -} \Big( \vec{\sigma}_{\vec{k}}^{+} \cdot \vec{\sigma}_{1} \Big) + C_{\vec{k}, +} \Big( \vec{\sigma}_{\vec{k}}^{-} \cdot \vec{\sigma}_{1} \Big)\bigg] \\
  \times a_{{^1}\!/{_2}, n_e, p}^{\dag} b_{-{^3}\!/{_2}, n_h, k_x-p}^{\dag} + \\
  + \bigg[ C_{\vec{k}, -} \Big( \vec{\sigma}_{\vec{k}}^{+} \cdot \vec{\sigma}_{-1} \Big) + C_{\vec{k}, +} \Big( \vec{\sigma}_{\vec{k}}^{-} \cdot \vec{\sigma}_{-1} \Big)\bigg]a_{-{^1}\!/{_2}, n_e, p}^{\dag} b_{{^3}\!/{_2}, n_h, k_x-p}^{\dag} \Bigg) + \\
  + P_{cv}^{\ast}(-k_y, p)\Phi^{\ast}\left( n_e, p; n_h, p+k_x; -k_y\right)
  \Bigg( \bigg[ C_{\vec{k}, -} \Big( \vec{\sigma}_{\vec{k}}^{+} \cdot \vec{\sigma}_{-1} \Big) + C_{\vec{k}, +} \Big( \vec{\sigma}_{\vec{k}}^{-} \cdot \vec{\sigma}_{-1} \Big)\bigg] \\
  \times b_{-{^3}\!/{_2}, n_h, -p-k_x} a_{{^1}\!/{_2}, n_e, p} + \\
  + \bigg[ C_{\vec{k}, -} \Big( \vec{\sigma}_{\vec{k}}^{+} \cdot \vec{\sigma}_{1} \Big) + C_{\vec{k}, +} \Big( \vec{\sigma}_{\vec{k}}^{-} \cdot \vec{\sigma}_{1} \Big)\bigg]b_{{^3}\!/{_2}, n_h, -p-k_x} a_{-{^1}\!/{_2}, n_e, p} \Bigg) + \\
  + P_{cv}(-k_y, p)\Phi\left( n_e, p; n_h, p+k_x; -k_y\right)
  \Bigg( \bigg[ \left(C_{\vec{k}, +}\right)^{\dag} \Big( \vec{\sigma}_{\vec{k}}^{+} \cdot \vec{\sigma}_{1} \Big) + \left(C_{\vec{k}, -}\right)^{\dag} \Big( \vec{\sigma}_{\vec{k}}^{-} \cdot \vec{\sigma}_{1} \Big)\bigg] \\ \times a_{{^1}\!/{_2}, n_e, p}^{\dag} b_{-{^3}\!/{_2}, n_h, -p-k_x}^{\dag} + \\
  + \bigg[ \left(C_{\vec{k}, +}\right)^{\dag} \Big( \vec{\sigma}_{\vec{k}}^{+} \cdot \vec{\sigma}_{-1} \Big) + \left(C_{\vec{k}, -}\right)^{\dag} \Big( \vec{\sigma}_{\vec{k}}^{-} \cdot \vec{\sigma}_{-1} \Big)\bigg]a_{-{^1}\!/{_2}, n_e, p}^{\dag} b_{{^3}\!/{_2}, n_h, -p-k_x}^{\dag} \Bigg) + \\
  + P_{cv}^{\ast}(k_y, p)\Phi^{\ast}\left( n_e, p; n_h, p-k_x; k_y\right)
  \Bigg( \bigg[ \left(C_{\vec{k}, +}\right)^{\dag} \Big( \vec{\sigma}_{\vec{k}}^{+} \cdot \vec{\sigma}_{-1} \Big) + \left(C_{\vec{k}, -}\right)^{\dag} \Big( \vec{\sigma}_{\vec{k}}^{-} \cdot \vec{\sigma}_{-1} \Big)\bigg] \\ \times b_{-{^3}\!/{_2}, n_h, k_x-p} a_{{^1}\!/{_2}, n_e, p} + \\
  + \bigg[ \left(C_{\vec{k}, +}\right)^{\dag} \Big( \vec{\sigma}_{\vec{k}}^{+} \cdot \vec{\sigma}_{1} \Big) + \left(C_{\vec{k}, -}\right)^{\dag} \Big( \vec{\sigma}_{\vec{k}}^{-} \cdot \vec{\sigma}_{1} \Big)\bigg] b_{{^3}\!/{_2}, n_h, k_x-p} a_{-{^1}\!/{_2}, n_e, p} \Bigg) \Bigg\}.
\end{array}
\end{equation}
\end{widetext}
Here volume $V$ of the 3D space can be represented as the product $V=SL_z$, where $L_z$ is the size of the 3D space in the direction $\vec{a}_3$. In the case of microcavity $L_z = L_c$. The matrix elements $P_{cv}$ of the band-to-band quantum transition were determined by the formula(A7) of Ref.[17]. In the case of conduction and valence bands of different parities it is assumed to be of allowed type according to the classification of Elliott [26-28] and does not depend on the wave vectors. The functions $\Phi\left(n_e, p; n_h, p-k_x; k_y\right)$ were determined by the formula (A11) of Ref.[17] and are also listed below
\begin{eqnarray}
 & P_{cv} (\vec{k}_{||}, g) = \frac{1}{v_0} {\displaystyle \intop_{v_0}} d\vec{\rho} U^{\ast}_{c,s,g} (\vec{\rho}) e^{ik_y \rho_y} (-i \hbar \frac{\partial}{\partial \rho}) U_{v,p,x,g-k_x}(\vec{\rho}); & \nonumber \\
 & \Phi\left(n_e, p; n_h, p-k_x; k_y\right) = e^{ik_ypl_0^2} \Phi\left(n_e, n_h; \vec{k}_{||}\right); & \nonumber \\
 & \Phi\left(n_e, n_h; \vec{k}_{||}\right) = {\displaystyle \intop_{- \infty}^{+\infty}} dy \varphi_{n_e}^{\ast}(y) \varphi_{n_h} \left( y+k_xl_0^2\right)e^{ik_yy}, & \label{phiortho}
\end{eqnarray}
where $\varphi_{n_e} (y)$ and $\varphi_{n_h} (y)$ are the real Landau quantization functions, whereas the functions $U_{c,s,g} (\rho)$ and $U_{v,p,x,g-k_x}(\rho)$ are the periodic parts of the electron Bloch functions in the conduction and valence bands. The last integral in the case $\vec{k}_{||} = 0$ is the normalization or orthogonality integral. The dipole active transitions ($\vec{k}_{||} = 0$) take place only in the case when $n_e=n_h$. It means that the 2D magnetoexciton can be created in the dipole-active transition only if it is constructed by the electron and hole on the Landau levels with the same quantum numbers $n_e=n_h$. In other words, the valence electron from the Landau level of quantization with a given number $n_v$ can be excited by light in the conduction band only on the level of Landau quantization with the same number $n_c=n_v$. It is true only for the dipole-active transitions. In the case of quadrupole-active transitions when the amplitudes of the quantum transitions are proportional to the projections of the wave vector $\vec{k}_{||}$ the selection rules are $n_e=n_h \pm 1$. Instead of the e-h pair representation in Ref.[17] the magnetoexciton creation operator was introduced. It depends on the wave vector $\vec{k}_{||}$, on the orbital momentum projection $M$ and on the Landau quantization numbers $n$ and $m$ as follows [21]
\begin{eqnarray}
&\Psi_{ex}^{\dag}(\vec{k}_{||},M,n,m) = \frac{1}{\sqrt{N}} \sum_{t}e^{ik_{y}tl_0^{2}} a_{s_z,n,\frac{k_{x}}{2}+t}^{\dag} b_{j_z,m,\frac{k_{x}}{2}-t}^{\dag}, & \nonumber \\
&s_{z} + j_{z} = M.&
\end{eqnarray}
The obtained Hamiltonian in the magnetoexciton representation looks as
\begin{widetext}
\begin{equation} \label{eq:hammagexph}
\begin{array}{c}
  \displaystyle \hat{H}_{magex-ph} = \left(-\frac{e}{m_0 l_0}\right)\sum_{\vec{k}(\vec{k}_{||}, k_{z})} \sum_{M = \pm 1} \sum_{n, m = 0}^{\infty} \sqrt{\frac{\hbar}{L_z \omega_{\vec{k}}}} \\
  \Bigg\{ P_{cv}(\vec{k}_{||})\Phi\left(n, m, \vec{k}_{||} \right) e^{\frac{ik_xk_{y}l_0^{2}}{2}}
  \bigg[ C_{\vec{k}, -} \Big( \vec{\sigma}_{\vec{k}}^{+} \cdot \vec{\sigma}_{M}^{\ast} \Big) + C_{\vec{k}, +} \Big( \vec{\sigma}_{\vec{k}}^{-} \cdot \vec{\sigma}_{M}^{\ast} \Big)\bigg] \\
  \times \hat{\Psi}_{ex}^{\dag}(\vec{k}_{||},M,n,m) + \\
  + P_{cv}^{\ast}(\vec{k}_{||})\Phi^{\ast}\left(n, m, \vec{k}_{||}\right) e^{\frac{-ik_xk_{y}l_0^{2}}{2}}
  \bigg[ \left(C_{\vec{k}, -}\right)^{\dag} \Big( \vec{\sigma}_{\vec{k}}^{+} \cdot \vec{\sigma}_{M}^{\ast} \Big)^{\ast} + \left(C_{\vec{k}, +}\right)^{\dag} \Big( \vec{\sigma}_{\vec{k}}^{-} \cdot \vec{\sigma}_{M}^{\ast} \Big)^{\ast} \bigg] \\
  \times \hat{\Psi}_{ex}(\vec{k}_{||},M,n,m) + \\
  + P_{cv}(-\vec{k}_{||})\Phi\left(n, m, -\vec{k}_{||}\right) e^{\frac{ik_xk_{y}l_0^{2}}{2}}
  \bigg[ \left(C_{\vec{k}, -}\right)^{\dag} \Big( \vec{\sigma}_{\vec{k}}^{+} \cdot \vec{\sigma}_{M}^{\ast} \Big)^{\ast} + \left(C_{\vec{k}, +}\right)^{\dag} \Big( \vec{\sigma}_{\vec{k}}^{-} \cdot \vec{\sigma}_{M}^{\ast} \Big)^{\ast} \bigg] \\
  \times \hat{\Psi}_{ex}^{\dag}(-\vec{k}_{||},-M,n,m) + \\
  + P_{cv}^{\ast}(-\vec{k}_{||}) \Phi^{\ast}\left(n, m, -\vec{k}_{||}\right) e^{\frac{-ik_xk_{y}l_0^{2}}{2}}
  \bigg[ C_{\vec{k}, -} \Big( \vec{\sigma}_{\vec{k}}^{+} \cdot \vec{\sigma}_{M}^{\ast} \Big) + C_{\vec{k}, +} \Big( \vec{\sigma}_{\vec{k}}^{-} \cdot \vec{\sigma}_{M}^{\ast} \Big)\bigg] \\
  \times \hat{\Psi}_{ex}(-\vec{k}_{||},-M,n,m) \Bigg\}.
\end{array}
\end{equation}
\end{widetext}
The interaction constant in the case of dipole-active transitions is proportional to $1/l_0$ and increases as $\sqrt{H}$ when the magnetic field strength $H$ increases. In the case of quadrupole-active transitions it does not depend on $H$, but is proportional to $|\vec{k}_{||}|$. The first two resonance terms describe the annihilation of the photon with circular polarization $\vec{\sigma}_{\vec{k}}^{\pm}$ and the creation of the magnetoexciton with the circular polarization $\vec{\sigma}_{M}$ and vice versa. The abilities of the photon to effectuate these transformations are determined by the scalar products $\left(\vec{\sigma}_{\vec{k}}^{\pm} \cdot \vec{\sigma}_{M}^{*}\right)$. The next two addenda are the anti-resonance terms describing the simultaneous creation or annihilation of the both partners namely of the photon and of the magnetoexciton with opposite sign 2D wave vectors $\vec{k}_{||}$ and $-\vec{k}_{||}$, and with opposite sign orbital momentum projections $M$ and $-M$.
	
Side by side with the electron-radiation interaction of the type $\sum\limits_{i}{\vec{A}({{r}_{i}})}{{\vec{\nabla }}_{i}}$ taken into account above there is also another interaction term proportional to the square of the vector potential $\vec{A} (\vec{r}_i)$ of the electromagnetic field in the form $\sum\limits_{i}{{{A}^{2}}({{r}_{i}})}$. It gives rise to a supplementary quadratic form in the photon operators containing the resonance and anti-resonance terms [27, 28]. They were neglected.
The interaction Hamiltonian was supplemented by the Hamiltonian $H_0$ of the free magnetoexcitons and photons
\begin{eqnarray}
H_{0} = {\sum_{\vec{k}_{||}} \sum_{M} \sum_{n,m}} E_{ex} (\vec{k}_{||},M,n,m)\times \nonumber \\
\times \hat{\Psi}_{ex}^{\dag}(\vec{k}_{||},M,n,m) \hat{\Psi}_{ex}(\vec{k}_{||},M,n,m) \nonumber \\
+ {\displaystyle \sum_{\vec{k} (\vec{k}_{||},k_z)}} \hbar \omega_{\vec{k}} \bigg[ \left(C_{\vec{k}, +}\right)^{\dag} C_{\vec{k}, +} + \left(C_{\vec{k}, -}\right)^{\dag} C_{\vec{k}, -} \bigg],
\end{eqnarray}
where $E_{ex} (\vec{k}_{||},M,n,m) = \hbar \omega_{ex} (\vec{k}_{||},M,n,m)$ is the energy of the 2D magnetoexciton. It contains the contributions of the cyclotron energies $n \hbar \omega_{ce} + m \hbar \omega_{ch}$ of the e-h pair forming the magnetoexciton and of the Coulomb e-h interaction in the presence of a strong magnetic field. The cyclotron frequencies $\omega_{ce}$ and $\omega_{ch}$ increase linearly as a function of $H$, whereas the Coulomb energy increases as a $\sqrt{H}$ in the same way as the constant of the magnetoexciton-photon interaction. We supposed that the Coulomb e-h interaction leading to the formation of the magnetoexciton is greater than the magnetoexciton-photon interaction leading to the formation of the magnetopolariton. It means that the ionization potential of the magnetoexciton $I_l = (e^2/\varepsilon l_0) \sqrt{\pi/2}$, where $\varepsilon$ is the dielectric constant, is greater than the Rabi energy $\hbar |\omega_R|$ introduced in the Ref.[21]. The magnetoexciton energy does not depend on $M$ when the Zeeman effect is not taken into account. The photon frequency depends on the 3D wave vector $\omega_{\vec{k}} = \frac{c}{n} \sqrt{\vec{k}_{||}^2 + k_{z}^2}$. The full Hamiltonian describing the magnetoexciton-polariton is
\begin{equation}
    H=H_{0}+H_{magex-ph}.
\end{equation}
These results will be generalized below taking into account the Rashba spin-orbit coupling, what means the use of the spinor-type wave functions (12) and (16) instead of the scalar ones [17, 21]. The Hamiltonian looks as
\begin{widetext}
\begin{eqnarray}
H_{e-rad}=(-\frac{e}{m_{0}})\sum_{\vec{k}(\vec{k}_{||}, k_{z})} \sqrt{\frac{2\pi \hbar}{V\omega_{\vec{k}}}} \sum_{i=1,2} \sum_{M_{v}=\pm 1} \sum_{\varepsilon=\varepsilon_{m},\varepsilon^{-}_{m}} \sum_{g,q}([C_{\vec{k},-}\vec{\sigma}^{+}_{\vec{k}}+C_{\vec{k},+}\vec{\sigma}^{-}_{\vec{k}}]\cdot  \nonumber \\
\cdot[\vec{P}(c,R_{i},g;v,M_{v},\varepsilon,q;\vec{k})a^{\dag}_{c,R_{i},g}a_{v,M_{v},\varepsilon,q}+\vec{P}(v,M_{v},\varepsilon,q;c,R_{i},g;\vec{k})a^{\dag}_{v,M_{v},\varepsilon,q}a_{c,R_{i},g}]+ \nonumber \\
+[(C_{\vec{k},-})^{\dag}\vec{\sigma}^{-}_{\vec{k}}+(C_{\vec{k},+})^{\dag}\vec{\sigma}^{+}_{\vec{k}}]\cdot[\vec{P}(c,R_{i},g;v,M_{v},\varepsilon,q;-\vec{k})\times \nonumber \\
\times a^{\dag}_{c,R_{i},g}a_{v,M_{v},\varepsilon,q}+\vec{P}(v,M_{v},\varepsilon,q;c,R_{i},g;-\vec{k})a^{\dag}_{v,M_{v},\varepsilon,q}a_{c,R_{i},g}])
\end{eqnarray}
\end{widetext}
The matrix elements will be discussed below. One of them has the form
\begin{eqnarray}
&\vec{P}(c,{{R}_{1}},g;v,{{M}_{v}},{{\varepsilon }^{-}}_{m},q;\vec{k})= &\nonumber \\
&=\int{{{d}^{2}}\vec{r}}\left\langle  {{{\hat{\psi }}}_{c,{{R}_{i}},g}}(\vec{r}) \right|{{e}^{i\vec{k}\vec{r}}}\hat{\vec{p}}\left| {{{\hat{\psi }}}_{v,{{M}_{v}},{{\varepsilon }^{-}}_{m},q}}(\vec{r}) \right\rangle = &\nonumber \\
&=\frac{a_{0}^{*}d_{m-3}^{-*}}{\sqrt{2}{{L}_{x}}}\int{{{d}^{2}}\vec{r}}U_{c,s,g}^{*}(\vec{r}){{e}^{-igx}}\varphi _{0}^{*}(y-g{{l}^{2}})\times & \nonumber \\
&\times {{e}^{i\vec{k}\vec{r}}}({{{\vec{a}}}_{1}}{{{\hat{p}}}_{x}}+{{{\vec{a}}}_{2}}{{{\hat{p}}}_{y}})({{U}_{v,p,x,q}}(\vec{r})\pm i{{U}_{v,p,y,q}}(\vec{r}))\times  &\nonumber \\
&\times {{e}^{iqx}}\varphi _{m-3}^{*}(y-p{{l}^{2}})-& \nonumber \\
&-\frac{b_{1}^{*}c_{m}^{-*}}{\sqrt{2}{{L}_{x}}}\int{{{d}^{2}}\vec{r}}U_{c,s,g}^{*}(\vec{r}){{e}^{-igx}}\times & \nonumber \\
&\times \varphi _{1}^{*}(y-g{{l}^{2}}){{e}^{i\vec{k}\vec{r}}}({{{\vec{a}}}_{1}}{{{\hat{p}}}_{x}}+{{{\vec{a}}}_{2}}{{{\hat{p}}}_{y}})\times  &\nonumber \\
&\times ({{U}_{v,p,x,q}}(\vec{r})\pm i{{U}_{v,p,y,q}}(\vec{r})){{e}^{iqx}}\varphi _{m}^{*}(y-p{{l}^{2}})&
\end{eqnarray}
One can represent the 2D coordinate vector $\vec{r}$ as a sum $\vec{r}=\vec{R}+\vec{\rho }$ of a lattice point vector $\vec{R}$ and of a small vector $\vec{\rho }$ changing inside the unit lattice cell with lattice constant ${{a}_{0}}$ and volume ${{v}_{0}}=a_{0}^{3}$. Any 2D semiconductor layer has at least the minimal width ${{a}_{0}}$ and the periodic parts ${{U}_{nk}}(\vec{r})$ are determined inside the elementary lattice cell. The periodic parts ${{U}_{nk}}(\vec{r})$ do not depend on $\vec{R}$ because ${{U}_{nk}}(\vec{R}+\vec{\rho })={{U}_{nk}}(\vec{\rho })$. From another side the envelope functions ${{\varphi }_{n}}(\vec{r})$ describing the Landau quantization have a spread of the order of magnetic length ${{l}_{0}}$ which is much greater than ${{a}_{0}}$ (${{l}_{0}}>>{{a}_{0}}$). It means that they practically do not depend on $\vec{\rho }$, i.e. ${{\varphi }_{n}}(\vec{R}+\vec{\rho })\cong {{\varphi }_{n}}(\vec{R})$. The matrix elements (27) contains some functions which do not depend on $\vec{R}$ and another ones, which do not depend on $\vec{\rho }$. Only the plane wave ${{e}^{i\vec{k}\vec{r}}}={{e}^{i\vec{k}\vec{R}+i\vec{k}\vec{\rho }}}$ contains both of them. The derivative $\frac{\partial }{\partial \vec{r}}$ acts in the same manner on the functions ${{U}_{nk}}(\vec{\rho })$ and ${{\varphi }_{n}}(\vec{R})$ because $\vec{R}$ and $\vec{\rho }$ are the components of $\vec{r}$. These properties suggested to transform the 2D integral on the variable $\vec{r}$ in two separate integrals on the variables $\vec{R}$ and $\vec{\rho }$ as follows
\begin{eqnarray}
\int{{{d}^{2}}\vec{r}}A(\vec{R})B(\vec{\rho })=\sum\limits_{{\vec{R}}}{A(\vec{R})}\int{{{d}^{2}}\vec{\rho }}B(\vec{\rho })= \nonumber \\
=\sum\limits_{{\vec{R}}}{A(\vec{R})}a_{0}^{2}\frac{1}{{{v}_{0}}}\int{{{d}^{3}}\vec{\rho }}B(\vec{\rho })= \nonumber \\
=\int{{{d}^{2}}\vec{R}}A(\vec{R})\frac{1}{{{v}_{0}}}\int\limits_{{{v}_{0}}}{d\vec{\rho }}B(\vec{\rho })
\end{eqnarray}
Here the small value $a_{0}^{2}$ is substituted by the infinitesimal differential value ${{d}^{2}}\vec{R}$ because $A(\vec{R})$ is a smooth function on $\vec{R}$. The integrals on the volume ${{v}_{0}}$ of the elementary lattice cell contain the quickly oscillating periodic parts $U_{c,s,g}^{{}}(\vec{\rho })$ and ${{U}_{v,p,i,q}}(\vec{\rho })$ belonging to s-type conduction band and to p-type valence band. They have different parities and obey to selection rules
\begin{eqnarray}
&\frac{1}{{{v}_{0}}}\int\limits_{{{v}_{0}}}{d\vec{\rho }}U_{c,s,g}^{*}(\vec{\rho }){{e}^{i{{k}_{y}}{{\rho }_{y}}}}{{U}_{v,p,i,q}}(\vec{\rho })=0, &\nonumber \\
&i,j=x,y,& \nonumber \\
&\frac{1}{{{v}_{0}}}\int\limits_{{{v}_{0}}}{d\vec{\rho }}U_{c,s,g}^{*}(\vec{\rho }){{e}^{i{{k}_{y}}{{\rho }_{y}}}}\frac{\partial }{\partial {{\rho }_{i}}}{{U}_{v,p,j,q}}(\vec{\rho })=0, &\nonumber \\
&\text{if }i\ne j.&
\end{eqnarray}
The case $i=j$ is different from zero and gives rise to the expression
\begin{equation}
\frac{1}{{{v}_{0}}}\int\limits_{{{v}_{0}}}{d\vec{\rho }}U_{c,s,g}^{*}(\vec{\rho }){{e}^{i{{k}_{y}}{{\rho }_{y}}}}\frac{\partial }{\partial {{\rho }_{i}}}{{U}_{v,pi,g-{{k}_{x}}}}(\vec{\rho })={{P}_{cv}}({{\vec{k}}_{||}},g)
\end{equation}
The last integral in the zeroth approximation is of the allowed type in the definition of Elliott [26, 27] and can be considered as a constant ${{P}_{cv}}({{\vec{k}}_{||}},g)\approx {{P}_{cv}}(0)$ which does not depend on the wave vectors ${{\vec{k}}_{||}}$ and $g$. Due to these selection rules the derivatives $\partial /\partial \vec{r}$ in the expression (27) must be taken only from the periodic parts ${{U}_{v,p,i,q}}(\vec{\rho })$ because all another integrals vanish.
The integration on the variable ${{R}_{x}}$ engages only the plane wave functions and gives rise to the selection rule for the 1D wave numbers $g,q,{{k}_{x}}$ as follows
\begin{equation}
\frac{1}{{{L}_{x}}}\int{{}}{{e}^{i{{R}_{x}}(q-g+{{k}_{x}})}}=\frac{2\pi }{{{L}_{x}}}\delta (q-g+{{k}_{x}})={{\delta }_{kr}}(q,g-{{k}_{x}})
\end{equation}
The integral on the variable ${{R}_{y}}$ engages only the Landau quantization functions ${{\varphi }_{n}}({{R}_{y}})$ and gives rise to the third selection rule concerning the numbers ${{n}_{e}}$ and ${{n}_{h}}$ of the Landau levels for electrons and holes. It looks as
\begin{eqnarray}
&\int\limits_{-\infty }^{\infty }{d{{R}_{y}}}\varphi _{{{n}_{e}}}^{*}({{R}_{y}}-g l_{0}^{2})\varphi _{{{n}_{h}}}^{*}({{R}_{y}}-(g-{{k}_{x}})l_{0}^{2}){{e}^{i{{k}_{y}}{{R}_{y}}}}= &\nonumber \\
&={{e}^{i{{k}_{y}}g l_{0}^{2}}}{{e}^{-i\frac{{{k}_{x}}{{k}_{y}}}{2}l_{0}^{2}}}\tilde{\phi }({{n}_{e}},{{n}_{h}};{{{\vec{k}}}_{||}})&
\end{eqnarray}
where
\begin{eqnarray}
\tilde{\phi }({{n}_{e}},{{n}_{h}};{{{\vec{k}}}_{||}})=\int\limits_{-\infty }^{\infty }{dy}\varphi _{{{n}_{e}}}^{{}}(y-\frac{{{k}_{x}}l_{0}^{2}}{2})\varphi _{{{n}_{h}}}^{{}}(y+\frac{{{k}_{x}}l_{0}^{2}}{2}){{e}^{i{{k}_{y}}y}}, \nonumber \\
{{e}^{-i\frac{{{k}_{x}}{{k}_{y}}}{2}l_{0}^{2}}}\tilde{\phi }({{n}_{e}},{{n}_{h}};{{{\vec{k}}}_{||}})=\phi ({{n}_{e}},{{n}_{h}};{{{\vec{k}}}_{||}}). \nonumber
\end{eqnarray}
The functions $\phi ({{n}_{e}},{{n}_{h}};{{\vec{k}}_{||}})$ (21) and $\tilde{\phi }({{n}_{e}},{{n}_{h}};{{\vec{k}}_{||}})$ (32) differ by the factor ${{e}^{-i\frac{{{k}_{x}}{{k}_{y}}}{2}l_{0}^{2}}}$.
Here we took into account that the Landau quantization functions ${{\varphi }_{n}}(y)$ are real.
This selection rule coincides with the formula (30) in the absence of the RSOC and its interpretation remains the same. Once again one can underlain that during the dipole-active band-to-band quantum transition the numbers of the Landau levels in the initial starting band as well as in the final arriving band coincide i.e. ${{n}_{e}}={{n}_{h}}$. It is true in the equal manner in the absence as well as in the presence of the RSOC.
Three separate integrations on $\vec{\rho },{{R}_{x}}\text{ and }{{R}_{y}}$ taking into account the selection rules (29), (31), (32) lead to the expression
\begin{eqnarray}
&\vec{P}(c,{{R}_{1}},g;v,{{M}_{v}},\varepsilon _{m}^{-},q;{{{\vec{k}}}_{||}})=& \nonumber \\
&={{\delta }_{kr}}(q,g-{{k}_{x}}){{{\vec{\sigma }}}_{{{M}_{v}}}}{{P}_{cv}}(0){{e}^{i{{k}_{y}}g l_{0}^{2}}}\times & \nonumber \\
&\times {{e}^{-i\frac{{{k}_{x}}{{k}_{y}}}{2}l_{0}^{2}}}[a_{0}^{*}d_{m-3}^{-*}\tilde{\phi }(0,m-3;{{{\vec{k}}}_{||}})-b_{1}^{*}c_{m}^{-*}\tilde{\phi }(1,m;{{{\vec{k}}}_{||}})], &\nonumber\\
& m\geq 3&
\end{eqnarray}
Here the vectors of the circular polarizations ${{\vec{\sigma }}_{{{M}_{v}}}}$ describing the valence electron states were introduced following the formula (17). One can introduce also the vectors of the heavy hole circular polarizations ${{\vec{\sigma }}_{{{M}_{h}}}}$ in the form
\begin{eqnarray}
{{{\vec{\sigma }}}_{{{M}_{v}}}}=\frac{1}{\sqrt{2}}({{{\vec{a}}}_{1}}\pm i{{{\vec{a}}}_{2}}),{{M}_{v}}=\pm 1, \nonumber \\
{{{\vec{\sigma }}}_{{{M}_{h}}}}={{{\vec{\sigma }}}^{*}}_{{{M}_{v}}}=\frac{1}{\sqrt{2}}({{{\vec{a}}}_{1}}\mp i{{{\vec{a}}}_{2}}),{{M}_{h}}=\mp 1
\end{eqnarray}
The magnetoexciton states are characterized by the quantum numbers ${{M}_{h}}$, ${{R}_{i}}$, $\varepsilon $ and by the wave vectors ${{\vec{k}}_{||}}$.
The general expressions for the matrix elements are
\begin{eqnarray}
&\vec{P}(c,{{R}_{i}},g;v,{{M}_{v}},\varepsilon ,q;{{{\vec{k}}}_{||}})=& \nonumber \\
&={{\delta }_{kr}}(q,g-{{k}_{x}}){{{\vec{\sigma }}}_{{{M}_{v}}}}{{P}_{cv}}(0){{e}^{i{{k}_{y}}g l_{0}^{2}}}T(c{{R}_{i}},\varepsilon ;{{{\vec{k}}}_{||}}){{e}^{-i\frac{{{k}_{x}}{{k}_{y}}}{2}l_{0}^{2}}},& \nonumber \\
&i=1,2,\text{ }{{\text{M}}_{v}}=\pm 1,\text{ }\varepsilon ={{\varepsilon }_{m}},& \nonumber \\
&\text{with }m=0,1,2,\text{and }\varepsilon =\varepsilon _{m}^{-},\text{with }m\ge 3&
\end{eqnarray}
The coefficients $T(c{{R}_{i}},\varepsilon ;{{\vec{k}}_{||}})$ have the forms
\begin{eqnarray}
&T({{R}_{1}},\varepsilon _{m}^{-};{{{\vec{k}}}_{||}})=[a_{0}^{*}d_{m-3}^{-*}\tilde{\phi }(0,m-3;{{{\vec{k}}}_{||}})-b_{1}^{*}c_{m}^{-*}\tilde{\phi }(1,m;{{{\vec{k}}}_{||}})],& \nonumber\\
&m\ge 3,&\nonumber \\
&T({{R}_{1}},{{\varepsilon }_{m}};{{{\vec{k}}}_{||}})=[-b_{1}^{*}\tilde{\phi }(1,m;{{{\vec{k}}}_{||}})], &\nonumber\\
&m=0,1,2, &\nonumber\\
&T({{R}_{2}},\varepsilon _{m}^{-};{{{\vec{k}}}_{||}})=[-c_{m}^{-*}\tilde{\phi }(0,m;{{{\vec{k}}}_{||}})],& \nonumber\\
&m\ge 3,&\nonumber \\
&T({{R}_{2}},\varepsilon _{m}^{{}};\vec{k})=[-\tilde{\phi }(0,m;{{{\vec{k}}}_{||}})], &\nonumber\\
&m=0,1,2&
\end{eqnarray}
Another matrix elements can be calculated in the similar way. They are
\begin{eqnarray}
&\vec{P}(v,{{M}_{v}},\varepsilon ,q;c,{{R}_{i}},g;-{{{\vec{k}}}_{||}})= &\nonumber \\
& ={{{\vec{P}}}^{*}}(c,{{R}_{i}},g;v,{{M}_{v}},\varepsilon ,q;{{{\vec{k}}}_{||}})= & \nonumber\\
& ={{\delta }_{kr}}(q,g-{{k}_{x}})\vec{\sigma }_{{{M}_{v}}}^{*}P_{cv}^{*}(0){{e}^{-i{{k}_{y}}g l_{0}^{2}}}{{e}^{i\frac{{{k}_{x}}{{k}_{y}}}{2}l_{0}^{2}}}{{T}^{*}}({{R}_{i}},\varepsilon ;{{{\vec{k}}}_{||}}), & \nonumber \\
& \vec{P}(c,{{R}_{i}},g;v,{{M}_{v}},\varepsilon ,q;-{{{\vec{k}}}_{||}})=& \nonumber \\
& -{{\delta }_{kr}}(q,g+{{k}_{x}})\vec{\sigma }_{{{M}_{v}}}^{{}}P_{cv}^{{}}(0){{e}^{-i{{k}_{y}}g l_{0}^{2}}}{{e}^{-i\frac{{{k}_{x}}{{k}_{y}}}{2}l_{0}^{2}}}T({{R}_{i}},\varepsilon ;-{{{\vec{k}}}_{||}}),& \nonumber  \\
& \vec{P}(v,{{M}_{v}},\varepsilon ,q;c,{{R}_{i}},g;{{{\vec{k}}}_{||}})= &  \\
& ={{\delta }_{kr}}(q,g+{{k}_{x}})\vec{\sigma }_{{{M}_{v}}}^{*}P_{cv}^{*}(0){{e}^{i{{k}_{y}}g l_{0}^{2}}}{{e}^{i\frac{{{k}_{x}}{{k}_{y}}}{2}l_{0}^{2}}}{{T}^{*}}({{R}_{i}},\varepsilon ;-{{{\vec{k}}}_{||}})& \nonumber
\end{eqnarray}
They permit to calculate the electron operator parts in the Hamiltonian (26) and to express them through the magnetoexciton creation and annihilation operators determined as follows
\begin{equation}
\hat{\psi }_{ex}^{\dagger }({{\vec{k}}_{||}},{{M}_{h}},{{R}_{i}},\varepsilon )=\frac{1}{\sqrt{N}}\sum\limits_{t}{{}}{{e}^{i{{k}_{y}}tl_{0}^{2}}}a_{{{R}_{i}},\frac{{{k}_{x}}}{2}+t}^{\dagger }b_{{{M}_{h}},\varepsilon ,\frac{{{k}_{x}}}{2}-t}^{\dagger }
\end{equation}
Here the electron and hole creation and annihilation operator where introduced
\begin{eqnarray}
& {{a}_{{{R}_{i}},g}}={{a}_{c,{{R}_{i}},g}}, & \nonumber \\
& {{a}_{v,{{M}_{v}},\varepsilon ,q}}=b_{-{{M}_{h}},\varepsilon ,-q}^{\dagger } &
\end{eqnarray}
Here we have supposed that the Coulomb electron-hole interaction leading to the formation of the magnetoexciton is greater than the magnetoexciton-photon interaction leading to the formation of the magnetopolariton. It means that the ionization potential of the magnetoexciton ${{I}_{l}}$ is greater than the Rabi energy $\hbar \left| {{\omega }_{R}} \right|\approx \frac{\left| e \right|}{{{m}_{0}}{{l}_{0}}}\left| {{P}_{cv}}(0) \right|\sqrt{\frac{\hbar }{{{L}_{z}}{{\omega }_{{\vec{k}}}}}}$. It was determined in Ref.[21].

The existence of the phase factors of the type ${{e}^{\pm i{{k}_{y}}gl_{0}^{2}}}$ in the expressions (35) and (37) similar with that entering in the definitions of the magnetoexciton creation operators permits to obtain the expressions
\begin{eqnarray}
&\sum\limits_{q,g}{{}}\vec{P}(c,{{R}_{i}},g;v,{{M}_{v}},\varepsilon ,q;{{{\vec{k}}}_{||}})a_{c,{{R}_{i}},g}^{\dagger }{{a}_{v,{{M}_{v}},\varepsilon ,q}}=& \nonumber \\
& =\vec{\sigma }_{{{M}_{h}}}^{*}{{P}_{cv}}(0)T({{R}_{i}},\varepsilon ;{{{\vec{k}}}_{||}})\sqrt{N}\hat{\psi }_{ex}^{\dagger }({{{\vec{k}}}_{||}},{{M}_{h}},{{R}_{i}},\varepsilon ),& \nonumber \\
& \sum\limits_{q,g}{{}}\vec{P}(v,{{M}_{v}},\varepsilon ,q;c,{{R}_{i}},g;-{{{\vec{k}}}_{||}})a_{v,{{M}_{v}},\varepsilon ,q}^{\dagger }{{a}_{c,{{R}_{i}},g}}=& \nonumber \\
& =\vec{\sigma }_{{{M}_{h}}}^{{}}{{P}^{*}}_{cv}(0){{T}^{*}}({{R}_{i}},\varepsilon ;{{{\vec{k}}}_{||}})\sqrt{N}\hat{\psi }_{ex}^{{}}({{{\vec{k}}}_{||}},{{M}_{h}},{{R}_{i}},\varepsilon ),& \nonumber \\
& \sum\limits_{q,g}{{}}\vec{P}(c,{{R}_{i}},g;v,{{M}_{v}},\varepsilon ,q;-{{{\vec{k}}}_{||}})a_{c,{{R}_{i}},g}^{\dagger }{{a}_{v,{{M}_{v}},\varepsilon ,q}}=& \nonumber \\
& =\vec{\sigma }_{{{M}_{h}}}^{*}{{P}_{cv}}(0)T({{R}_{i}},\varepsilon ;-{{{\vec{k}}}_{||}})\sqrt{N}\hat{\psi }_{ex}^{\dagger }(-{{{\vec{k}}}_{||}},{{M}_{h}},{{R}_{i}},\varepsilon ),& \nonumber \\
& \sum\limits_{q,g}{{}}\vec{P}(v,{{M}_{v}},\varepsilon ,q;c,{{R}_{i}},g;{{{\vec{k}}}_{||}})a_{v,{{M}_{v}},\varepsilon ,q}^{\dagger }{{a}_{c,{{R}_{i}},g}}=& \\
& =\vec{\sigma }_{{{M}_{h}}}^{{}}{{P}^{*}}_{cv}(0){{T}^{*}}({{R}_{i}},\varepsilon ;-{{{\vec{k}}}_{||}})\sqrt{N}\hat{\psi }_{ex}^{{}}(-{{{\vec{k}}}_{||}},{{M}_{h}},{{R}_{i}},\varepsilon ),& \nonumber
\end{eqnarray}
In the Ref.[21] the Hamiltonian of the electron-radiation interaction was deduced in the absence of the RSOC. In its presence the mentioned Hamiltonian also can be expressed in compact form through the photon and magnetoexciton creation and annihilation operators. As earlier we introduced the values $N=S/2\pi l_{0}^{2}$, $V=S{{L}_{z}}$, where ${{L}_{z}}$ is the size of the 3D space in direction perpendicular to the layer. In the case of microcavity ${{L}_{z}}$ equals to the cavity length ${{L}_{c}}$. The electron-radiation interaction has the form
\begin{widetext}
\begin{eqnarray}
{{{\hat{H}}}_{e-rad}}=\left( -\frac{e}{{{m}_{0}}{{l}_{0}}} \right)\sum\limits_{\vec{k}({{{\vec{k}}}_{||}},{{k}_{z}})}{\sum\limits_{{{M}_{h}}=\pm 1}{\sum\limits_{i=1,2}^{{}}{\sum\limits_{\varepsilon ={{\varepsilon }_{m}},\varepsilon _{m}^{-}}{\sqrt{\frac{\hbar }{{{L}_{z}}{{\omega }_{{\vec{k}}}}}}}}}}\times  \nonumber\\
\times \{{{P}_{cv}}(0)T\left( {{R}_{i}},\varepsilon ,{{{\vec{k}}}_{||}} \right)[{{C}_{\vec{k},-}}(\vec{\sigma }_{{\vec{k}}}^{+}\cdot \vec{\sigma }_{{{M}_{h}}}^{*})+{{C}_{\vec{k},+}}(\vec{\sigma }_{{\vec{k}}}^{-}\cdot \vec{\sigma }_{{{M}_{h}}}^{*})]\hat{\psi }_{ex}^{\dagger }({{{\vec{k}}}_{||}},{{M}_{h}},{{R}_{i}},\varepsilon )+ \nonumber\\
+{{P}^{*}}_{cv}(0){{T}^{*}}\left( {{R}_{i}},\varepsilon ,{{{\vec{k}}}_{||}} \right)[{{\left( {{C}_{\vec{k},-}} \right)}^{\dagger }}(\vec{\sigma }_{{\vec{k}}}^{-}\cdot \vec{\sigma }_{{{M}_{h}}}^{{}})+{{\left( {{C}_{\vec{k},+}} \right)}^{\dagger }}(\vec{\sigma }_{{\vec{k}}}^{+}\cdot \vec{\sigma }_{{{M}_{h}}}^{{}})]\hat{\psi }_{ex}^{{}}({{{\vec{k}}}_{||}},{{M}_{h}},{{R}_{i}},\varepsilon )+ \nonumber\\
+{{P}_{cv}}(0)T\left( {{R}_{i}},\varepsilon ,-{{{\vec{k}}}_{||}} \right)[{{\left( {{C}_{\vec{k},-}} \right)}^{\dagger }}(\vec{\sigma }_{{\vec{k}}}^{-}\cdot \vec{\sigma }_{{{M}_{h}}}^{*})+{{\left( {{C}_{\vec{k},+}} \right)}^{\dagger }}(\vec{\sigma }_{{\vec{k}}}^{+}\cdot \vec{\sigma }_{{{M}_{h}}}^{*})]\hat{\psi }_{ex}^{\dagger }(-{{{\vec{k}}}_{||}},{{M}_{h}},{{R}_{i}},\varepsilon )+ \nonumber\\
+{{P}^{*}}_{cv}(0){{T}^{*}}\left( {{R}_{i}},\varepsilon ,-{{{\vec{k}}}_{||}} \right)[{{C}_{\vec{k},-}}(\vec{\sigma }_{{\vec{k}}}^{+}\cdot \vec{\sigma }_{{{M}_{h}}}^{{}})+{{C}_{\vec{k},+}}(\vec{\sigma }_{{\vec{k}}}^{-}\cdot \vec{\sigma }_{{{M}_{h}}}^{{}})]\hat{\psi }_{ex}^{{}}(-{{{\vec{k}}}_{||}},{{M}_{h}},{{R}_{i}},\varepsilon )\}
\end{eqnarray}
\end{widetext}
This expression is similar with the Hamiltonian (23) in the absence of the RSOC. The difference is related with the more complicate coefficients $T({{R}_{i}},\varepsilon ;{{\vec{k}}_{||}})$ in comparison with their components $\phi (n,m;{{\vec{k}}_{||}})$ and with more large number of heavy hole states included in the last variant. Now the Coulomb interaction between charged carriers in the presence of the RSOC will be investigated.
\section{The Coulomb interaction in the 2D electron-hole system under the influence of the Rashba spin-orbit coupling}
The Coulomb interaction in the 2D e-h system taking into account the Rashba spin-orbit coupling was discussed in Ref. [19, 22]. Below we will remember these results including all valence electron states (12). In the present description the multi-component electron field contains a larger variety of the valence band states in comparison with [19, 22]. For the very beginning the properties of the density operator of the electron field $\hat{\rho }(\vec{r})$ and of its Fourier components $\hat{\rho }(\vec{Q})$ will be discussed. To this end the Fermi-type creation and annihilation operators of the electron on different states were introduced. They are denoted as $a_{{{R}_{i}},g}^{\dagger },a_{{{R}_{i}},g}^{{}}$ for the conduction band Rashba-type states (16) $\left| {{\psi }_{c}}({{R}_{i}},g;r) \right\rangle $, as $a_{{{M}_{v}},{{\varepsilon }_{m}},g}^{\dagger },a_{{{M}_{v}},{{\varepsilon }_{m}},g}^{{}}$ for the spinor valence band states (12) $\left| {{\psi }_{v}}({{M}_{v}},{{\varepsilon }_{m}},g;r) \right\rangle $ and as $a_{{{M}_{v}},\varepsilon _{m}^{-},g}^{\dagger },a_{{{M}_{v}},\varepsilon _{m}^{-},g}^{{}}$ for another spinor valence band states (12) $\left| {{\psi }_{v}}({{M}_{v}},\varepsilon _{m}^{-},g;r) \right\rangle $. These spinor-type functions have a form of a column with two components corresponding to two spin projections on the direction of the magnetic field. The conjugate functions $\left\langle  {{\psi }_{c}}({{R}_{i}},g;r) \right|$, $\left\langle  {{\psi }_{c}}({{R}_{i}},g;r) \right|$ and $\left\langle  {{\psi }_{v}}({{M}_{v}},\varepsilon _{m}^{-},g;r) \right|$ have a form of a row with two components conjugate to the components of the columns. With the aid of the electron creation and annihilation operators and of the spinor-type wave functions the creation and annihilation operators ${{\hat{\Psi }}^{\dagger }}(r)$ and $\hat{\Psi }(r)$ of the multi-component electron field can be written as
\begin{eqnarray}
\hat{\Psi }(r)=\sum\limits_{i=1,2}{\sum\limits_{g}{{}}}\left| {{\psi }_{c}}({{R}_{i}},g;r) \right\rangle {{a}_{{{R}_{i}},g}}+ \nonumber \\
+\sum\limits_{{{M}_{v}}}{\sum\limits_{{{\varepsilon }_{m}}}{\sum\limits_{g}{{}}}}\left| {{\psi }_{v}}({{M}_{v}},{{\varepsilon }_{m}},g;r) \right\rangle {{a}_{{{M}_{v}},{{\varepsilon }_{m}},g}}+ \nonumber\\
+\sum\limits_{{{M}_{v}}}{\sum\limits_{\varepsilon _{m}^{-}}{\sum\limits_{g}{{}}}}\left| {{\psi }_{v}}({{M}_{v}},\varepsilon _{m}^{-},g;r) \right\rangle {{a}_{{{M}_{v}},\varepsilon _{m}^{-},g}}, \nonumber\\
{{{\hat{\Psi }}}^{\dagger }}(r)=\sum\limits_{i=1,2}{\sum\limits_{g}{{}}}\left\langle  {{\psi }_{c}}({{R}_{i}},g;r) \right|a_{{{R}_{i}},g}^{\dagger }+ \nonumber\\
+\sum\limits_{{{M}_{v}}}{\sum\limits_{{{\varepsilon }_{m}}}{\sum\limits_{g}{{}}}}\left\langle  {{\psi }_{v}}({{M}_{v}},{{\varepsilon }_{m}},g;r) \right|a_{{{M}_{v}},{{\varepsilon }_{m}},g}^{\dagger }+ \nonumber\\
+\sum\limits_{{{M}_{v}}}{\sum\limits_{\varepsilon _{m}^{-}}{\sum\limits_{g}{{}}}}\left\langle  {{\psi }_{v}}({{M}_{v}},\varepsilon _{m}^{-},g;r) \right|a_{{{M}_{v}},\varepsilon _{m}^{-},g}^{\dagger }
\end{eqnarray}
The density operator of the electron field $\hat{\rho }(\vec{r})$ and its Fourier components $\hat{\rho }(\vec{Q})$ are determined by the expressions
\begin{eqnarray}
\hat{\rho }(\vec{r})={{{\hat{\Psi }}}^{\dagger }}(r)\hat{\Psi }(r), \nonumber \\
\hat{\rho }(\vec{Q})=\int{{{d}^{2}}\vec{r}}\hat{\rho }(\vec{r}){{e}^{i\vec{Q}\vec{r}}}
\end{eqnarray}
The density operator looks as
\begin{widetext}
\begin{eqnarray}
\hat{\rho }(\vec{r})=\sum\limits_{i,j=1,2}{\sum\limits_{g,q}{{}}}a_{{{R}_{j}},q}^{\dagger }{{a}_{{{R}_{i}},g}}\left\langle {{\psi }_{c}}({{R}_{j}},q;r)|{{\psi }_{c}}({{R}_{i}},g;r) \right\rangle + \nonumber\\
+\sum\limits_{{{M}_{v}},{{{{M}'}}_{v}}}{\sum\limits_{{{\varepsilon }_{m}},{{\varepsilon }_{{{m}'}}}}{\sum\limits_{g,q}{{}}}}a_{{{{{M}'}}_{v}},{{\varepsilon }_{{{m}'}}},q}^{\dagger }{{a}_{{{M}_{v}},{{\varepsilon }_{m}},g}}\left\langle {{\psi }_{v}}({{{{M}'}}_{v}},{{\varepsilon }_{{{m}'}}},q;r) \right.\left| {{\psi }_{v}}({{M}_{v}},{{\varepsilon }_{m}},g;r) \right\rangle + \nonumber\\
+\sum\limits_{{{M}_{v}},{{{{M}'}}_{v}}}{\sum\limits_{\varepsilon _{m}^{-},\varepsilon _{{{m}'}}^{-}}{\sum\limits_{g,q}{{}}a_{{{{{M}'}}_{v}},\varepsilon _{{{m}'}}^{-},q}^{\dagger }{{a}_{{{M}_{v}},\varepsilon _{m}^{-},g}}}}\left\langle {{\psi }_{v}}({{{{M}'}}_{v}},\varepsilon _{{{m}'}}^{-},q;r) \right.\left| {{\psi }_{v}}({{M}_{v}},\varepsilon _{m}^{-},g;r) \right\rangle + \nonumber\\
+\sum\limits_{{{M}_{v}},{{{{M}'}}_{v}}}{\sum\limits_{\varepsilon _{{{m}'}}^{{}},\varepsilon _{m}^{-}}{\sum\limits_{g,q}{{}}a_{{{{{M}'}}_{v}},\varepsilon _{{{m}'}}^{{}},q}^{\dagger }{{a}_{{{M}_{v}},\varepsilon _{m}^{-},g}}}}\left\langle {{\psi }_{v}}({{{{M}'}}_{v}},\varepsilon _{{{m}'}}^{{}},q;r) \right.\left| {{\psi }_{v}}({{M}_{v}},\varepsilon _{m}^{-},g;r) \right\rangle + \nonumber\\
+\sum\limits_{{{M}_{v}},{{{{M}'}}_{v}}}{\sum\limits_{\varepsilon _{{{m}'}}^{-},\varepsilon _{m}^{{}}}{\sum\limits_{g,q}{{}}a_{{{{{M}'}}_{v}},\varepsilon _{{{m}'}}^{-},q}^{\dagger }{{a}_{{{M}_{v}},\varepsilon _{m}^{{}},g}}}}\left\langle {{\psi }_{v}}({{{{M}'}}_{v}},\varepsilon _{{{m}'}}^{-},q;r) \right.\left| {{\psi }_{v}}({{M}_{v}},\varepsilon _{m}^{{}},g;r) \right\rangle + \nonumber\\
+\sum\limits_{i=1,2}{\sum\limits_{{{M}_{v}}}{\sum\limits_{{{\varepsilon }_{m}}}{\sum\limits_{g,q}{{}}}}}a_{{{R}_{i}},q}^{\dagger }{{a}_{{{M}_{v}},{{\varepsilon }_{m}},g}}\left\langle {{\psi }_{c}}({{R}_{i}},q;r)|{{\psi }_{v}}({{M}_{v}},{{\varepsilon }_{m}},g;r) \right\rangle + \nonumber\\
+\sum\limits_{i=1,2}{\sum\limits_{{{M}_{v}}}{\sum\limits_{\varepsilon _{m}^{-}}{\sum\limits_{g,q}{{}}}}}a_{{{R}_{i}},q}^{\dagger }{{a}_{{{M}_{v}},\varepsilon _{m}^{-},g}}\left\langle {{\psi }_{c}}({{R}_{i}},q;r)|{{\psi }_{v}}({{M}_{v}},\varepsilon _{m}^{-},g;r) \right\rangle + \nonumber\\
+\sum\limits_{i=1,2}{\sum\limits_{{{M}_{v}}}{\sum\limits_{{{\varepsilon }_{m}}}{\sum\limits_{g,q}{{}}}}}a_{{{M}_{v}},{{\varepsilon }_{m}},q}^{\dagger }{{a}_{{{R}_{i}},g}}\left\langle {{\psi }_{v}}({{M}_{v}},{{\varepsilon }_{m}},q;r)|{{\psi }_{c}}({{R}_{i}},g;r) \right\rangle + \nonumber\\
+\sum\limits_{i=1,2}{\sum\limits_{{{M}_{v}}}{\sum\limits_{\varepsilon _{m}^{-}}{\sum\limits_{g,q}{{}}}}}a_{{{M}_{v}},\varepsilon _{m}^{-},q}^{\dagger }{{a}_{{{R}_{i}},g}}\left\langle {{\psi }_{v}}({{M}_{v}},\varepsilon _{m}^{-},q;r)|{{\psi }_{c}}({{R}_{i}},g;r) \right\rangle
\end{eqnarray}
\end{widetext}
The Fourier components $\hat{\rho }(\vec{Q})$ of the density operator determine the Coulomb interaction between the electrons. They will be calculated below taking into account the spinor-type wave functions (12) and (16). For example, the first term in the expressions (44) looks as
\begin{eqnarray}
&{{{\hat{\rho }}}_{c-c}}({{R}_{1}};{{R}_{1}};\vec{Q})=& \nonumber \\
& =\sum\limits_{q,g}{a_{{{R}_{1}},q}^{\dagger }{{a}_{{{R}_{1}},g}}}\left\langle  {{\psi }_{c}}({{R}_{1}},q;\vec{r}) \right|\left. {{\psi }_{c}}({{R}_{1}},g;\vec{r}) \right\rangle =& \nonumber \\
& =\sum\limits_{q,g}{a_{{{R}_{1}},q}^{\dagger }{{a}_{{{R}_{1}},g}}}\int{{{d}^{2}}\vec{r}}U_{c,s,q}^{*}(\vec{r}){{U}_{c,s,g}}(\vec{r})\frac{{{e}^{i(g+{{Q}_{x}}-q)x}}}{{{L}_{x}}}\times & \nonumber \\
& \times [{{\left| {{a}_{0}} \right|}^{2}}\varphi _{0}^{*}(y-q l_{0}^{2})\varphi _{0}^{{}}(y-g l_{0}^{2})+ & \nonumber \\
& +{{\left| {{b}_{1}} \right|}^{2}}\varphi _{1}^{*}(y-q l_{0}^{2})\varphi _{1}^{{}}(y-gl_{0}^{2})], & \nonumber \\
 & \vec{r}=\vec{R}+\vec{\rho }&
\end{eqnarray}
Following the formula (28) it is necessary to separate the integration of the quickly varying periodic parts on the volume ${{v}_{0}}$ of the elementary lattice cell and the integration of the slowly varying envelope parts on the lattice point vectors $\vec{R}$ as follows
\begin{eqnarray}
&\frac{1}{{{v}_{0}}}\int\limits_{{{v}_{0}}}{d\rho }U_{c,s,g+{{Q}_{x}}}^{*}(\rho )U_{c,s,g}^{{}}(\rho ){{e}^{i{{Q}_{y}}{{\rho }_{y}}}}=1+O(\vec{Q}),& \nonumber \\
& \frac{1}{{{L}_{x}}}\int{{{e}^{i(g-q+{{Q}_{x}}){{R}_{x}}}}d{{R}_{x}}}={{\delta }_{kr}}(q,g+{{Q}_{x}}),& \nonumber \\
& \tilde{\phi }(n,m;\vec{Q})= & \nonumber\\
& =\int{d{{R}_{y}}}\varphi _{n}^{*}\left( {{R}_{y}}-\frac{{{Q}_{x}}l_{0}^{2}}{2} \right)\varphi _{m}^{{}}\left( {{R}_{y}}+\frac{{{Q}_{x}}l_{0}^{2}}{2} \right){{e}^{i{{Q}_{y}}{{R}_{y}}}}= & \nonumber \\
 & ={{{\tilde{\phi }}}^{*}}(m,n;-\vec{Q})&
\end{eqnarray}
Here $O(\vec{Q})$ is an infinitesimal value much smaller than unity, tending to zero in the limit $Q\to 0$. It will be neglected in all calculations below. The calculation give rise to the final form
\begin{eqnarray}
&{{{\hat{\rho }}}_{c-c}}({{R}_{1}};{{R}_{1}};\vec{Q})=[{{\left| {{a}_{0}} \right|}^{2}}\tilde{\phi }(0,0;\vec{Q})+& \nonumber \\
 & +{{\left| {{b}_{1}} \right|}^{2}}\tilde{\phi }(1,1;\vec{Q})]\hat{\rho }({{R}_{1}};{{R}_{1}};\vec{Q})= & \nonumber \\
 & =\tilde{S}({{R}_{1}};{{R}_{1}};\vec{Q})\hat{\rho }({{R}_{1}};{{R}_{1}};\vec{Q}),&  \nonumber \\
 & \hat{\rho }({{R}_{1}};{{R}_{1}};\vec{Q})=\sum\limits_{t}{{}}{{e}^{i{{Q}_{y}}tl_{0}^{2}}}a_{{{R}_{1}},t+\frac{{{Q}_{x}}}{2}}^{\dagger }a_{{{R}_{1}},t-\frac{{{Q}_{x}}}{2}}^{{}}, & \nonumber \\
 & \tilde{S}({{R}_{1}};{{R}_{1}};\vec{Q})=[{{\left| {{a}_{0}} \right|}^{2}}\tilde{\phi }(0,0;\vec{Q})+{{\left| {{b}_{1}} \right|}^{2}}\tilde{\phi }(1,1;\vec{Q})]&
\end{eqnarray}
The expression (47) looks as a product of one numeral factor $\tilde{S}({{R}_{1}};{{R}_{1}};\vec{Q})$, which concerns the concrete electron spinor state and another operator type factor of the general form
\begin{eqnarray}
\hat{\rho }(\xi ,\eta ;\vec{Q})= \nonumber \\
=\sum\limits_{t}{{}}{{e}^{i{{Q}_{y}}t l_{0}^{2}}}a_{\xi ,t+\frac{{{Q}_{x}}}{2}}^{\dagger }{{a}_{\eta ,t-\frac{{{Q}_{x}}}{2}}}= \nonumber \\
={{{\hat{\rho }}}^{\dagger }}(\eta ,\xi ;-\vec{Q})
\end{eqnarray}
It will be met in all expressions listed below, but with different meanings of $\xi $ and $\eta $, as follows
\begin{eqnarray}
&{{{\hat{\rho }}}_{c-c}}({{R}_{2}};{{R}_{2}};\vec{Q})=\tilde{S}({{R}_{2}};{{R}_{2}};\vec{Q})\hat{\rho }({{R}_{2}};{{R}_{2}};\vec{Q}),& \nonumber \\
& \tilde{S}({{R}_{2}};{{R}_{2}};\vec{Q})=\tilde{\phi }(0,0;\vec{Q}),& \nonumber \\
& {{{\hat{\rho }}}_{c-c}}({{R}_{1}};{{R}_{2}};\vec{Q})=\tilde{S}({{R}_{1}};{{R}_{2}};\vec{Q})\hat{\rho }({{R}_{1}};{{R}_{2}};\vec{Q}),& \nonumber \\
& \tilde{S}({{R}_{1}};{{R}_{2}};\vec{Q})=b_{1}^{*}\tilde{\phi }(1,0;\vec{Q}),& \nonumber \\
& {{{\hat{\rho }}}_{c-c}}({{R}_{2}};{{R}_{1}};\vec{Q})=\tilde{S}({{R}_{2}};{{R}_{1}};\vec{Q})\hat{\rho }({{R}_{2}};{{R}_{1}};\vec{Q})& \nonumber \\
& =\rho _{c-c}^{\dagger }({{R}_{1}};{{R}_{2}};-\vec{Q}),& \nonumber \\
& \tilde{S}({{R}_{2}};{{R}_{1}};\vec{Q})=b_{1}^{{}}\tilde{\phi }(0,1;\vec{Q}) &
\end{eqnarray}
One of the valence electron density fluctuation operator looks as
\begin{eqnarray}
&{{{\hat{\rho }}}_{v-v}}({{M}_{v}},\varepsilon _{m}^{-},{{{{M}'}}_{v}},\varepsilon _{{{m}'}}^{-};\vec{Q})= & \nonumber \\
 & \sum\limits_{q,g}{{}}a_{{{M}_{v}},\varepsilon _{m}^{-},q}^{\dagger }{{a}_{{{{{M}'}}_{v}},\varepsilon _{{{m}'}}^{-},g}}\int{{{d}^{2}}\vec{r}}{{e}^{i\vec{Q}\vec{r}}}\times & \nonumber \\
 & \times \left\langle  {{\psi }_{v}}({{M}_{v}},\varepsilon _{m}^{-},q;\vec{r}) \right|\left. {{\psi }_{v}}({{{{M}'}}_{v}},\varepsilon _{{{m}'}}^{-},g;\vec{r}) \right\rangle = & \nonumber \\
 & =\tilde{S}({{M}_{v}},\varepsilon _{m}^{-};{{{{M}'}}_{v}},\varepsilon _{{{m}'}}^{-};\vec{Q})\hat{\rho }({{M}_{v}},\varepsilon _{m}^{-};{{M}_{v}},\varepsilon _{{{m}'}}^{-}\vec{Q}), & \nonumber \\
 & \tilde{S}({{M}_{v}},\varepsilon _{m}^{-};{{{{M}'}}_{v}},\varepsilon _{{{m}'}}^{-};\vec{Q})= & \nonumber \\
 & ={{\delta }_{{{M}_{v}},{{{{M}'}}_{v}}}}[d_{m-3}^{-}d_{{m}'-3}^{-*}\tilde{\phi }(m-3,{m}'-3;\vec{Q})+ & \nonumber \\
 & +c_{m}^{-}c_{{{m}'}}^{-*}\tilde{\phi }(m,{m}';\vec{Q})], & \nonumber \\
 & m,{m}'\ge 3&
\end{eqnarray}
Here we have taken into account the following property of the valence band periodic parts
\begin{eqnarray}
&\frac{1}{{{v}_{0}}}\int\limits_{{{v}_{0}}}{d\vec{\rho }}U_{v,p,i,g+{{Q}_{x}}}^{*}(\rho )U_{v,p,j,g}^{{}}(\rho ){{e}^{i{{Q}_{y}}{{\rho }_{y}}}}=& \nonumber \\
 & ={{\delta }_{ij}}+O(\vec{Q}),& \nonumber \\
 & i,j=x,y&
\end{eqnarray}
They leads to the Kronekker symbol ${{\delta }_{{{M}_{v}},{{{{M}'}}_{v}}}}$ in the expression (50) and in the next ones concerning the valence band, as follows
\begin{eqnarray}
&{{{\hat{\rho }}}_{v-v}}({{M}_{v}},\varepsilon _{m}^{{}};{{{{M}'}}_{v}},\varepsilon _{{{m}'}}^{{}};\vec{Q})=& \nonumber \\
 & =\tilde{S}({{M}_{v}},\varepsilon _{m}^{{}};{{{{M}'}}_{v}},\varepsilon _{{{m}'}}^{{}};\vec{Q})\hat{\rho }({{M}_{v}},\varepsilon _{m}^{{}};{{M}_{v}},\varepsilon _{{{m}'}}^{{}};\vec{Q}),& \nonumber \\
 & \tilde{S}({{M}_{v}},\varepsilon _{m}^{{}};{{{{M}'}}_{v}},\varepsilon _{{{m}'}}^{{}};\vec{Q})={{\delta }_{{{M}_{v}},{{{{M}'}}_{v}}}}\tilde{\phi }(m,{m}';\vec{Q}), & \nonumber \\
 & m,{m}'=0,1,2,& \nonumber \\
 & {{{\hat{\rho }}}_{v-v}}({{{{M}'}}_{v}},\varepsilon _{{{m}'}}^{{}};{{M}_{v}},\varepsilon _{m}^{-};\vec{Q})= & \nonumber \\
 & =\tilde{S}({{{{M}'}}_{v}},\varepsilon _{{{m}'}}^{{}};{{M}_{v}},\varepsilon _{m}^{-};\vec{Q})\hat{\rho }({{M}_{v}},\varepsilon _{{{m}'}}^{{}};{{M}_{v}},\varepsilon _{m}^{-};\vec{Q}), & \nonumber \\
 & \tilde{S}({{{{M}'}}_{v}},\varepsilon _{{{m}'}}^{{}};{{M}_{v}},\varepsilon _{m}^{-};\vec{Q})={{\delta }_{{{M}_{v}},{{{{M}'}}_{v}}}}c_{m}^{-*}\tilde{\phi }({m}',m;\vec{Q}),& \nonumber \\
 & {m}'=0,1,2,\text{ }m\ge 3,& \nonumber \\
 & {{{\hat{\rho }}}_{v-v}}({{M}_{v}},\varepsilon _{m}^{-};{{{{M}'}}_{v}},\varepsilon _{{{m}'}}^{{}};\vec{Q})= & \nonumber\\
 & =\tilde{S}({{M}_{v}},\varepsilon _{m}^{-};{{{{M}'}}_{v}},\varepsilon _{{{m}'}}^{{}};\vec{Q})\hat{\rho }({{M}_{v}},\varepsilon _{m}^{-};{{M}_{v}},\varepsilon _{{{m}'}}^{{}};\vec{Q}), & \nonumber\\
 & \tilde{S}({{M}_{v}},\varepsilon _{m}^{-};{{{{M}'}}_{v}},\varepsilon _{{{m}'}}^{{}};\vec{Q})={{\delta }_{{{M}_{v}},{{{{M}'}}_{v}}}}c_{m}^{-}\tilde{\phi }(m,{m}';\vec{Q}), & \nonumber\\
 & m\ge 3,\text{ }{m}'=0,1,2&
\end{eqnarray}
As usual they obey to the equalities
\begin{eqnarray}
\hat{\rho }_{v-v}^{\dagger }(\xi ;\eta ;\vec{Q})=\hat{\rho }_{v-v}^{{}}(\eta ;\xi ;-\vec{Q}), \nonumber\\
\hat{\rho }(\xi ;\eta ;\vec{Q})=\hat{\rho }_{{}}^{\dagger }(\eta ;\xi ;-\vec{Q}), \nonumber \\
\tilde{\phi }(n;m;\vec{Q})={{{\tilde{\phi }}}^{*}}(m;n;-\vec{Q}), \nonumber \\
\tilde{S}(\xi ;\eta ;\vec{Q})={{{\tilde{S}}}^{*}}(\eta ;\xi ;-\vec{Q})
\end{eqnarray}
Up till now we have deal with the intraband density operators ${{\hat{\rho }}_{c-c}}(\xi ;\eta ;\vec{Q})$ and ${{\hat{\rho }}_{v-v}}(\xi ;\eta ;\vec{Q})$. The interband density operators ${{\hat{\rho }}_{c-v}}(\xi ;\eta ;\vec{Q})$ and ${{\hat{\rho }}_{v-c}}(\xi ;\eta ;\vec{Q})$ depend on the interband exchange electron densities of the type $U_{c,s,g+{{Q}_{x}}}^{*}(\rho )\frac{1}{\sqrt{2}}({{U}_{v,p,x,g}}(\rho )\pm i{{U}_{v,p,y,g}}(\rho ))$ and its complex conjugate value. They contain the quickly oscillating periodic parts with different parities and the ortoganality integral on the elementary lattice cell has an infinitesimal value
\begin{eqnarray}
O(\vec{Q})=\frac{1}{{{v}_{0}}}\int\limits_{{{v}_{0}}}{d\rho }U_{c,s,g+{{Q}_{x}}}^{*}(\rho ){{e}^{i{{Q}_{y}}{{\rho }_{y}}}}\times  \nonumber \\
\times \frac{1}{\sqrt{2}}({{U}_{v,p,x,g}}(\rho )\pm i{{U}_{v,p,y,g}}(\rho ))
\end{eqnarray}
 This integral is different from zero if one takes into account, for example, the term $i{{Q}_{y}}{{\rho }_{y}}$ appearing in the series expansion of the function ${{e}^{i{{Q}_{y}}{{\rho }_{y}}}}$. It gives rise to the interband dipole momentum ${{\vec{d}}_{cv}}$ with the component
 \begin{eqnarray}
 {{d}_{cv,y}}=\frac{e}{{{v}_{0}}}\int\limits_{{{v}_{0}}}{d\rho }U_{c,s,g}^{*}(\rho ){{\rho }_{y}}{{U}_{v,p,y,g}}(\rho ), \nonumber \\
  O(\vec{Q})\approx {{Q}_{y}}{{d}_{cv,y}}
 \end{eqnarray}
The Coulomb interaction depending on the interband exchange electron densities ${{\hat{\rho }}_{cv}}(\vec{Q}){{\hat{\rho }}_{vc}}(-\vec{Q})$ has a form of the dipole-dipole interaction instead of the charge-charge interaction, which takes place only in the intraband cases. It is known as a long-range Coulomb interaction and gives rise to the longitudinal-transverse splitting of the three-fold degenerate levels of the dipole-active excitons in the cubic crystals [27]. Such type effects with the participation of the 2D magnetoexcitons were not investigated up till now, to the best of our knowledge, and remain outside the present review article.

The density operator $\hat{\rho }(\vec{Q})$ in the frame of the electron spinor states (12) and (16) looks as
\begin{eqnarray}
\hat{\rho }(\vec{Q})=\sum\limits_{i,j}{{}}{{{\hat{\rho }}}_{c-c}}({{R}_{i}};{{R}_{j}};\vec{Q})+ \nonumber\\
+\sum\limits_{{{M}_{v}},{{{{M}'}}_{v}}}{\sum\limits_{{{\varepsilon }_{m}},{{\varepsilon }_{{{m}'}}}}{{}}}{{{\hat{\rho }}}_{v-v}}({{M}_{v}},{{\varepsilon }_{m}};{{{{M}'}}_{v}},{{\varepsilon }_{{{m}'}}};\vec{Q})+ \nonumber\\
+\sum\limits_{{{M}_{v}},{{{{M}'}}_{v}}}{\sum\limits_{\varepsilon _{m}^{-},\varepsilon _{{{m}'}}^{-}}{{}}}{{{\hat{\rho }}}_{v-v}}({{M}_{v}},\varepsilon _{m}^{-};{{{{M}'}}_{v}},\varepsilon _{{{m}'}}^{-};\vec{Q})+ \nonumber\\
+\sum\limits_{{{M}_{v}},{{{{M}'}}_{v}}}{\sum\limits_{\varepsilon _{m}^{-},\varepsilon _{{{m}'}}^{{}}}{{}}}{{{\hat{\rho }}}_{v-v}}({{{{M}'}}_{v}},\varepsilon _{{{m}'}}^{{}};{{M}_{v}},\varepsilon _{m}^{-};\vec{Q})+ \nonumber\\
+\sum\limits_{{{M}_{v}},{{{{M}'}}_{v}}}{\sum\limits_{\varepsilon _{m}^{-},\varepsilon _{{{m}'}}^{{}}}{{}}}{{{\hat{\rho }}}_{v-v}}({{M}_{v}},\varepsilon _{m}^{-};{{{{M}'}}_{v}},\varepsilon _{{{m}'}}^{{}};\vec{Q})+ \nonumber\\
+\sum\limits_{i}{\sum\limits_{{{M}_{v}}}{\sum\limits_{{{\varepsilon }_{m}}}{{}}}}{{{\hat{\rho }}}_{c-v}}({{R}_{i}};{{M}_{v}},{{\varepsilon }_{m}};\vec{Q})+ \nonumber\\
+\sum\limits_{i}{\sum\limits_{{{M}_{v}}}{\sum\limits_{\varepsilon _{m}^{-}}{{}}}}{{{\hat{\rho }}}_{c-v}}({{R}_{i}};{{M}_{v}},\varepsilon _{m}^{-};\vec{Q})+ \nonumber\\
+\sum\limits_{i}{\sum\limits_{{{M}_{v}}}{\sum\limits_{{{\varepsilon }_{m}}}{{}}}}{{{\hat{\rho }}}_{v-c}}({{M}_{v}},{{\varepsilon }_{m}};{{R}_{i}};\vec{Q})+ \nonumber\\
+\sum\limits_{i}{\sum\limits_{{{M}_{v}}}{\sum\limits_{\varepsilon _{m}^{-}}{{}}}}{{{\hat{\rho }}}_{v-c}}({{M}_{v}},\varepsilon _{m}^{-};{{R}_{i}};\vec{Q})
\end{eqnarray}
The first five terms of this expression depend on the intraband electron densities and determine the charge-charge Coulomb interaction. The last four terms depend on the interband electron densities and lead to the dipole-dipole long-range Coulomb interaction.

The strength of the Coulomb interaction is determined by coefficients ${{a}_{n}},{{b}_{n}},{{c}_{n}},{{d}_{n}}$ of the spinor-type wave function (12) and (16) as well as by the normalization and orthogonality-type integrals $\tilde{\phi }(n,m,\vec{Q})$. They have the properties:
\begin{eqnarray}
\tilde{\phi }(n,m,\vec{Q})={{e}^{-\frac{{{Q}^{2}}{{l}^{2}}}{4}}}{{A}_{n,m}}(\vec{Q}) \nonumber\\
{{A}_{n,m}}(0)={{\delta }_{n,m}}
\end{eqnarray}
The diagonal coefficients ${{A}_{n,n}}(\vec{Q})$ with $n=0,1,3$ will be calculated below. The nondiagonal coefficients with $n\ne m$ in the limit $Q\to 0$ are proportional to the vector components ${{Q}_{i}}$ in a degree $|n-m|$. They can be neglected in the zeroth order approximation together with another corrections denoted as $O(\vec{Q})$. It essentially deminish the number of the actual components of the density operator $\hat{\rho }(\vec{Q})$.

In the zeroth order approximation, neglecting the corrections of the order $O(\vec{Q})$ we will deal only with diagonal terms which permit the simplified denotations
\begin{eqnarray}
\hat{\rho }(\xi ;\xi ;\vec{Q})=\hat{\rho }(\xi ;\vec{Q}), \nonumber \\
\tilde{S}(\xi ;\xi ;\vec{Q})=\tilde{S}(\xi ;\vec{Q})={{e}^{-\frac{{{Q}^{2}}l_{0}^{2}}{4}}}S(\xi ;\vec{Q})
\end{eqnarray}
The concrete values of the coefficients $S(\xi ;\vec{Q})$ are
\begin{eqnarray}
& S({{R}_{1}};\vec{Q})=[|{{a}_{0}}{{|}^{2}}{{A}_{0,0}}(\vec{Q})+|{{b}_{1}}{{|}^{2}}{{A}_{1,1}}(\vec{Q})],& \nonumber \\
& S({{R}_{2}};\vec{Q})={{A}_{0,0}}(\vec{Q}), & \nonumber\\
& S({{\varepsilon }_{m}};\vec{Q})={{A}_{m,m}}(\vec{Q}),m=0,1,2,& \nonumber \\
& S(\varepsilon _{m}^{-};\vec{Q})= & \nonumber\\
& =[|d_{m-3}^{-}{{|}^{2}}{{A}_{m-3,m-3}}(\vec{Q})+|c_{m}^{-}{{|}^{2}}{{A}_{m,m}}(\vec{Q})],& \nonumber \\
& m\ge 3&
\end{eqnarray}
The calculated values ${{A}_{m,m}}(\vec{Q})$ equal to
\begin{eqnarray}
& {{A}_{0,0}}(\vec{Q})=1,\text{ }{{A}_{1,1}}(\vec{Q})=\left( 1-\frac{{{Q}^{2}}l_{0}^{2}}{2} \right), & \nonumber \\
& {{A}_{3,3}}(\vec{Q})=1-\frac{3}{2}{{Q}^{2}}l_{0}^{2}+\frac{3}{8}{{Q}^{4}}l_{0}^{4}-\frac{1}{48}{{Q}^{6}}l_{0}^{6} &
\end{eqnarray}
The diagonal part of the density operator $\hat{\rho }(\vec{Q})$ looks as
\begin{eqnarray}
& \hat{\rho }(\vec{Q})={{e}^{-\frac{{{Q}^{2}}l_{0}^{2}}{4}}}\{\sum\limits_{i}{{}}S({{R}_{i}};\vec{Q})\hat{\rho }({{R}_{i}};\vec{Q})+ & \nonumber \\
& +\sum\limits_{{{M}_{v}}}{\sum\limits_{{{\varepsilon }_{m}}}{{}}S({{M}_{v}},{{\varepsilon }_{m}};\vec{Q})}\hat{\rho }({{M}_{v}},{{\varepsilon }_{m}};\vec{Q})+ & \nonumber \\
& +\sum\limits_{{{M}_{v}}}{\sum\limits_{\varepsilon _{m}^{-}}{{}}S({{M}_{v}},\varepsilon _{m}^{-};\vec{Q})}\hat{\rho }({{M}_{v}},\varepsilon _{m}^{-};\vec{Q})\}&
\end{eqnarray}
It contains two separate contributions from the conduction and valence bands. The latter contribution in its turn can be represented as due to the electrons of the full filled valence band extracting the contribution of the holes created in its frame. To show it one can introduce the hole creation and annihilation operators as follows
\begin{eqnarray}
b_{{{M}_{h}},\varepsilon ,q}^{\dagger }={{a}_{v,-{{M}_{v}},\varepsilon ,-q}}, \nonumber \\
{{b}_{{{M}_{h}},\varepsilon ,q}}=a_{v,-{{M}_{v}},\varepsilon ,-q}^{\dagger }, \nonumber \\
\varepsilon ={{\varepsilon }_{m}},m=0,1,2, \nonumber \\
\varepsilon =\varepsilon _{m}^{-},m\ge 3
\end{eqnarray}
It leads to the relation
\begin{eqnarray}
& {{{\hat{\rho }}}_{v}}(-{{M}_{v}},\varepsilon ,\vec{Q})=N{{\delta }_{kr}}(\vec{Q},0)-{{{\hat{\rho }}}_{h}}({{M}_{h}},\varepsilon ,\vec{Q}), & \nonumber \\
& N=\frac{S}{2\pi l_{0}^{2}}&
\end{eqnarray}
where the hole density operator ${{\rho }_{h}}({{M}_{h}},\varepsilon ,\vec{Q})$ looks as
\begin{equation}
{{\hat{\rho }}_{h}}({{M}_{h}},\varepsilon ,\vec{Q})=\sum\limits_{t}{{e}^{-i{{Q}_{y}}tl_{0}^{2}}}b_{{{M}_{h}},\varepsilon ,t+\frac{{{Q}_{x}}}{2}}^{\dagger }b_{{{M}_{h}},\varepsilon ,t-\frac{{{Q}_{x}}}{2}}
\end{equation}
The constant part $N{{\delta }_{kr}}(\vec{Q},0)$in (63) created by electron of the full filled valence band is compensated by the influence of the positive electric charges of the background nuclei. In the jelly model of the system their presence is taken into account excluding from the Hamiltonian of the Coulomb interaction the point $\vec{Q}=0$ [29]. Taking into account the fully neutral system of the bare electrons and of the positive jelly background we will operate only with the conduction band electrons and with the holes in the valence band. In this electron-hole description the density operator $\hat{\rho }(\vec{Q})$ becomes equal
\begin{eqnarray}
\hat{\rho }(\vec{Q})={{{\hat{\rho }}}_{e}}(\vec{Q})-{{{\hat{\rho }}}_{h}}(\vec{Q}), \nonumber \\
\vec{Q}\ne 0
\end{eqnarray}
where
\begin{eqnarray}
& {{{\hat{\rho }}}_{h}}(\vec{Q})=\sum\limits_{{{M}_{h}},{{\varepsilon }_{m}}}{{}}\tilde{S}({{\varepsilon }_{m}};\vec{Q}){{{\hat{\rho }}}_{h}}({{M}_{h}},{{\varepsilon }_{m}};\vec{Q})+& \nonumber \\
& +\sum\limits_{{{M}_{h}},\varepsilon _{m}^{-}}{{}}\tilde{S}(\varepsilon _{m}^{-};\vec{Q}){{{\hat{\rho }}}_{h}}({{M}_{h}},\varepsilon _{m}^{-};\vec{Q})= & \nonumber \\
& ={{e}^{-\frac{{{Q}^{2}}l_{0}^{2}}{4}}}\{\sum\limits_{{{M}_{h}},{{\varepsilon }_{m}}}{{}}S({{\varepsilon }_{m}};\vec{Q}){{{\hat{\rho }}}_{h}}({{M}_{h}},{{\varepsilon }_{m}};\vec{Q})+ & \nonumber \\
& +\sum\limits_{{{M}_{h}},\varepsilon _{m}^{-}}{{}}S(\varepsilon _{m}^{-};\vec{Q}){{{\hat{\rho }}}_{h}}({{M}_{h}},\varepsilon _{m}^{-};\vec{Q})\}, & \nonumber \\
& {{{\hat{\rho }}}_{e}}(\vec{Q})={{{\hat{\rho }}}_{c}}(\vec{Q})= & \nonumber \\
& =\sum\limits_{i=1,2}{{}}\tilde{S}({{R}_{i}};\vec{Q}){{{\hat{\rho }}}_{e}}({{R}_{i}};\vec{Q})= & \nonumber \\
& ={{e}^{-\frac{{{Q}^{2}}l_{0}^{2}}{4}}}\sum\limits_{i=1,2}{{}}S({{R}_{i}};\vec{Q}){{{\hat{\rho }}}_{e}}({{R}_{i}};\vec{Q}) &
\end{eqnarray}
The Hamiltonian of the Coulomb interaction of the initial bare electrons can be expressed through the electron field and density operators (42) and (56), as follows
\begin{eqnarray}
& {{H}_{Coul}}= & \nonumber \\
& =\frac{1}{2}\int{d\vec{1}}\int{d\vec{2}}{{{\hat{\Psi }}}^{\dagger }}(1){{{\hat{\Psi }}}^{\dagger }}(2)V(1-2)\hat{\Psi }(2)\hat{\Psi }(1)= & \nonumber \\
& =\frac{1}{2}\sum\limits_{{\vec{Q}}}{{}}V(\vec{Q})\int{d\vec{1}}\int{d\vec{2}}{{e}^{i\vec{Q}(\vec{1}-\vec{2})}}{{{\hat{\Psi }}}^{\dagger }}(1)\hat{\rho }(2)\hat{\Psi }(1)= & \nonumber \\
& =\frac{1}{2}\sum\limits_{{\vec{Q}}}{{}}V(\vec{Q})\int{\int{d\vec{r}}}{{e}^{i\vec{Q}\vec{r}}}{{{\hat{\Psi }}}^{\dagger }}(\vec{r})\hat{\rho }(-\vec{Q})\hat{\Psi }(\vec{r}), & \nonumber \\
& V(\vec{Q})=\frac{2\pi {{e}^{2}}}{{{\varepsilon }_{0}}S|\vec{Q}|}&
\end{eqnarray}
$V(\vec{Q})$ is the Fourier transform of the Coulomb interaction of the electrons situated on the surface of the 2D layer with the area S and dielectric constant ${{\varepsilon }_{0}}$ of the medium. The expression ${{\hat{\Psi }}^{\dagger }}(\vec{r})\hat{\rho }(-\vec{Q})\hat{\Psi }(\vec{r})$ contains the density operator $\hat{\rho }(-\vec{Q})$ intercalated between the field operators ${{\hat{\Psi }}^{\dagger }}(\vec{r})$ and $\hat{\Psi }(\vec{r})$. The operator $\hat{\Psi }(\vec{r})$ cannot be transposed over the operator $\hat{\rho }(-\vec{Q})$ because they do not commute, but its nonoperator part expressed through the spinor-type wave function can be transposed forming together with the conjugate wave function of the field operator ${{\hat{\Psi }}^{\dagger }}(\vec{r})$ a scalar.
After the integration on the coordinate $\vec{r}$ the quadratic intercalated density operators will appear with the forms
\begin{eqnarray}
& K(\xi ;\eta ;x;\eta ;\vec{Q})=& \\
& =\sum\limits_{t}{{}}{{e}^{i{{Q}_{y}}t l_{0}^{2}}}a_{\xi ,t+\frac{{{Q}_{x}}}{2}}^{\dagger }\hat{\rho }(x;y;-\vec{Q})a_{\eta ,t-\frac{{{Q}_{x}}}{2}}^{{}}= & \nonumber \\
& =\sum\limits_{t}{\sum\limits_{s}{{}}}{{e}^{i{{Q}_{\eta }}t l_{0}^{2}}}{{e}^{-i{{Q}_{y}}s l_{0}^{2}}}a_{\xi ,t+\frac{{{Q}_{x}}}{2}}^{\dagger }a_{x,s-\frac{{{Q}_{x}}}{2}}^{\dagger }a_{y,s+\frac{{{Q}_{x}}}{2}}^{{}}a_{\eta ,t-\frac{{{Q}_{x}}}{2}}^{{}}= & \nonumber \\
& =\hat{\rho }(\xi ;\eta ;\vec{Q})\hat{\rho }(x;y;-\vec{Q})-{{\delta }_{\eta ,x}}\hat{\rho }(\xi ;y;0) & \nonumber
\end{eqnarray}
The same relations remain in the electron-hole description.
The commutation relations between the density operators are the followings
\begin{eqnarray}
\hat{\rho }(\xi ;\eta ;\vec{Q})=\sum\limits_{t}{{e}^{i{{Q}_{y}}t l_{0}^{2}}}a_{\xi ,t+\frac{{{Q}_{x}}}{2}}^{\dagger }a_{\eta ,t-\frac{{{Q}_{x}}}{2}}, \nonumber\\
\hat{\rho }(x;y;\vec{P})=\sum\limits_{t}{{}}{{e}^{i{{P}_{y}}t l_{0}^{2}}}a_{x,t+\frac{{{P}_{x}}}{2}}^{\dagger }a_{y,t-\frac{{{P}_{x}}}{2}},\nonumber \\
\left[\hat{\rho }(\xi ;\eta ;\vec{Q}),\hat{\rho }(x;y;\vec{P})\right]= \nonumber \\
={{\delta }_{x,\eta }}\hat{\rho }(\xi ;y;\vec{P}+\vec{Q}){{e}^{\frac{i{{(\vec{P}\times \vec{Q})}_{z}}l_{0}^{2}}{2}}}- \nonumber \\
-{{\delta }_{\xi ,y}}\hat{\rho }(x;\eta ;\vec{P}+\vec{Q}){{e}^{\frac{-i{{(\vec{P}\times \vec{Q})}_{z}}l_{0}^{2}}{2}}}= \nonumber \\
=\cos \left( \frac{{{[\vec{P}\times \vec{Q}]}_{z}}l_{0}^{2}}{2} \right)[{{\delta }_{x,\eta }}\hat{\rho }(\xi ;y;\vec{P}+\vec{Q})- \nonumber \\
-{{\delta }_{\xi ,y}}\hat{\rho }(x;\eta ;\vec{P}+\vec{Q})]+ \nonumber \\
+i\sin \left( \frac{{{(\vec{P}\times \vec{Q})}_{z}}l_{0}^{2}}{2} \right)[{{\delta }_{x,\eta }}\hat{\rho }(\xi ;y;\vec{P}+\vec{Q})+ \nonumber \\
+{{\delta }_{\xi ,y}}\hat{\rho }(x;\eta ;\vec{P}+\vec{Q})]
\end{eqnarray}
The factor ${{e}^{-\frac{{{Q}^{2}}l_{0}^{2}}{4}}}$ arising from the product of the density operators $\hat{\rho }(\vec{Q})$ and $\hat{\rho }(-\vec{Q})$ being multiplied with the coefficient $V(\vec{Q})$ gives rise to the coefficient $W(\vec{Q})$ describing the effective Coulomb interaction in the conditions of the Landau quantization
\begin{equation}
W(\vec{Q})=V(\vec{Q}){{e}^{-\frac{{{Q}^{2}}l_{0}^{2}}{2}}}
\end{equation}
Excluding the intercalations, the Hamiltonian of the Coulomb interaction in the presence of the Landau quantization and Rashba spin-orbit coupling has the form:
\begin{widetext}
\begin{eqnarray}
&{{H}_{Coul}}=\frac{1}{2}\sum\limits_{{\vec{Q}}}{{}}W(\vec{Q})\{\sum\limits_{i,j}{{}}S({{R}_{i}};\vec{Q})S({{R}_{j}};-\vec{Q})[\hat{\rho }({{R}_{i}};\vec{Q})\hat{\rho }({{R}_{j}};-\vec{Q})-{{\delta }_{i,j}}\hat{\rho }({{R}_{i}};0)]+& \nonumber \\
&+\sum\limits_{{{M}_{v}},{{{{M}'}}_{v}}}{\sum\limits_{{{\varepsilon }_{m}},\varepsilon _{{{m}'}}^{{}}}{{}}}S({{M}_{v}},{{\varepsilon }_{m}};\vec{Q})S({{{{M}'}}_{v}},{{\varepsilon }_{{{m}'}}};-\vec{Q})[\hat{\rho }({{M}_{v}},{{\varepsilon }_{m}};\vec{Q})\hat{\rho }({{{{M}'}}_{v}},{{\varepsilon }_{{{m}'}}};-\vec{Q})-{{\delta }_{{{M}_{v}},{{{{M}'}}_{v}}}}{{\delta }_{m,{m}'}}\hat{\rho }({{M}_{v}},{{\varepsilon }_{m}};0)]+& \nonumber \\
&+\sum\limits_{{{M}_{v}},{{{{M}'}}_{v}}}{\sum\limits_{\varepsilon _{m}^{-},\varepsilon _{{{m}'}}^{-}}{{}}}S({{M}_{v}},\varepsilon _{m}^{-};\vec{Q})S({{{{M}'}}_{v}},\varepsilon _{{{m}'}}^{-};-\vec{Q})[\hat{\rho }({{M}_{v}},\varepsilon _{m}^{-};\vec{Q})\hat{\rho }({{{{M}'}}_{v}},\varepsilon _{{{m}'}}^{-};-\vec{Q})-{{\delta }_{{{M}_{v}},{{{{M}'}}_{v}}}}{{\delta }_{m,{m}'}}\hat{\rho }({{M}_{v}},\varepsilon _{m}^{-};0)]+ &\nonumber \\
&+\sum\limits_{i=1,2}{\sum\limits_{{{M}_{v}},{{\varepsilon }_{m}}}{{}}}S({{R}_{i}};\vec{Q})S({{M}_{v}},{{\varepsilon }_{m}};-\vec{Q})\hat{\rho }({{R}_{i}};\vec{Q})\hat{\rho }({{M}_{v}},{{\varepsilon }_{m}};-\vec{Q})+ &\nonumber \\
&+\sum\limits_{i=1,2}{\sum\limits_{{{M}_{v}},\varepsilon _{m}^{-}}{{}}}S({{R}_{i}};\vec{Q})S({{M}_{v}},\varepsilon _{m}^{-};-\vec{Q})\hat{\rho }({{R}_{i}};\vec{Q})\hat{\rho }({{M}_{v}},\varepsilon _{m}^{-};-\vec{Q})+ &\nonumber \\
&+\sum\limits_{i=1,2}{\sum\limits_{{{M}_{v}},{{\varepsilon }_{m}}}{{}}}S({{M}_{v}},{{\varepsilon }_{m}};\vec{Q})S({{R}_{i}};-\vec{Q})\hat{\rho }({{M}_{v}},{{\varepsilon }_{m}};\vec{Q})\hat{\rho }({{R}_{i}};-\vec{Q})+ &\nonumber \\
&+\sum\limits_{i=1,2}{\sum\limits_{{{M}_{v}},\varepsilon _{m}^{-}}{{}}}S({{M}_{v}},\varepsilon _{m}^{-};\vec{Q})S({{R}_{i}};-\vec{Q})\hat{\rho }({{M}_{v}},\varepsilon _{m}^{-};\vec{Q})\hat{\rho }({{R}_{i}};-\vec{Q})+ &\nonumber \\
&+\sum\limits_{{{M}_{v}},{{{{M}'}}_{v}}}{\sum\limits_{\varepsilon _{m}^{{}},\varepsilon _{{{m}'}}^{-}}{{}}}S({{M}_{v}},{{\varepsilon }_{m}};\vec{Q})S({{{{M}'}}_{v}},\varepsilon _{{{m}'}}^{-};-\vec{Q})\hat{\rho }({{M}_{v}},{{\varepsilon }_{m}};\vec{Q})\hat{\rho }({{{{M}'}}_{v}},\varepsilon _{{{m}'}}^{-};-\vec{Q})+ &\nonumber \\
&+\sum\limits_{{{M}_{v}},{{{{M}'}}_{v}}}{\sum\limits_{\varepsilon _{m}^{-},\varepsilon _{{{m}'}}^{{}}}{{}}}S({{M}_{v}},\varepsilon _{m}^{-};\vec{Q})S({{{{M}'}}_{v}},{{\varepsilon }_{{{m}'}}};-\vec{Q})\hat{\rho }({{M}_{v}},\varepsilon _{m}^{-};\vec{Q})\hat{\rho }({{{{M}'}}_{v}},{{\varepsilon }_{{{m}'}}};-\vec{Q})\}&
\end{eqnarray}
The Hamiltonian of the Coulomb interaction in the electron-hole representation looks as
\begin{eqnarray}
&{{H}_{Coul}}=\frac{1}{2}\sum\limits_{{\vec{Q}}}{{}}W(\vec{Q})\{\sum\limits_{i,j}{{}}S({{R}_{i}};\vec{Q})S({{R}_{j}};-\vec{Q})[{{{\hat{\rho }}}_{e}}({{R}_{i}};\vec{Q}){{{\hat{\rho }}}_{e}}({{R}_{j}};-\vec{Q})-{{\delta }_{i,j}}{{{\hat{\rho }}}_{e}}({{R}_{i}};0)]+ &\nonumber \\
&+\sum\limits_{{{M}_{h}},{{{{M}'}}_{h}}}{\sum\limits_{{{\varepsilon }_{m}},\varepsilon _{{{m}'}}^{{}}}{{}}}S({{M}_{h}},{{\varepsilon }_{m}};\vec{Q})S({{{{M}'}}_{h}},{{\varepsilon }_{{{m}'}}};-\vec{Q})[{{{\hat{\rho }}}_{h}}({{M}_{h}},{{\varepsilon }_{m}};\vec{Q}){{{\hat{\rho }}}_{h}}({{{{M}'}}_{h}},{{\varepsilon }_{{{m}'}}};-\vec{Q})-{{\delta }_{{{M}_{h}},{{{{M}'}}_{h}}}}{{\delta }_{m,{m}'}}{{{\hat{\rho }}}_{h}}({{M}_{h}},{{\varepsilon }_{m}};0)]+ &\nonumber \\
&+\sum\limits_{{{M}_{h}},{{{{M}'}}_{h}}}{\sum\limits_{\varepsilon _{m}^{-},\varepsilon _{{{m}'}}^{-}}{{}}}S({{M}_{h}},\varepsilon _{m}^{-};\vec{Q})S({{{{M}'}}_{h}},\varepsilon _{{{m}'}}^{-};-\vec{Q})[{{{\hat{\rho }}}_{h}}({{M}_{h}},\varepsilon _{m}^{-};\vec{Q}){{{\hat{\rho }}}_{h}}({{{{M}'}}_{h}},\varepsilon _{{{m}'}}^{-};-\vec{Q})-{{\delta }_{{{M}_{h}},{{{{M}'}}_{h}}}}{{\delta }_{m,{m}'}}{{{\hat{\rho }}}_{h}}({{M}_{h}},\varepsilon _{m}^{-};0)]+& \nonumber \\
&+\sum\limits_{{{M}_{h}},{{{{M}'}}_{h}}}{\sum\limits_{\varepsilon _{m}^{{}},\varepsilon _{{{m}'}}^{-}}{{}}}S({{M}_{h}},{{\varepsilon }_{m}};\vec{Q})S({{{{M}'}}_{h}},\varepsilon _{{{m}'}}^{-};-\vec{Q}){{{\hat{\rho }}}_{h}}({{M}_{h}},{{\varepsilon }_{m}};\vec{Q}){{{\hat{\rho }}}_{h}}({{{{M}'}}_{h}},\varepsilon _{{{m}'}}^{-};-\vec{Q})+& \nonumber \\
&+\sum\limits_{{{M}_{h}},{{{{M}'}}_{h}}}{\sum\limits_{\varepsilon _{m}^{-},\varepsilon _{{{m}'}}^{{}}}{{}}}S({{M}_{h}},\varepsilon _{m}^{-};\vec{Q})S({{{{M}'}}_{h}},{{\varepsilon }_{{{m}'}}};-\vec{Q}){{{\hat{\rho }}}_{h}}({{M}_{h}},\varepsilon _{m}^{-};\vec{Q}){{{\hat{\rho }}}_{h}}({{{{M}'}}_{h}},{{\varepsilon }_{{{m}'}}};-\vec{Q})- & \nonumber \\
&-\sum\limits_{i}{\sum\limits_{{{M}_{h}},{{\varepsilon }_{m}}}{{}}}S({{R}_{i}};\vec{Q})S({{M}_{h}},{{\varepsilon }_{m}};-\vec{Q}){{{\hat{\rho }}}_{e}}({{R}_{i}};\vec{Q}){{{\hat{\rho }}}_{h}}({{M}_{h}},{{\varepsilon }_{m}};-\vec{Q})- & \nonumber \\
&-\sum\limits_{i}{\sum\limits_{{{M}_{h}},\varepsilon _{m}^{-}}{{}}}S({{R}_{i}};\vec{Q})S({{M}_{h}},\varepsilon _{m}^{-};-\vec{Q}){{{\hat{\rho }}}_{e}}({{R}_{i}};\vec{Q}){{{\hat{\rho }}}_{h}}({{M}_{h}},\varepsilon _{m}^{-};-\vec{Q})- &\nonumber \\
&-\sum\limits_{i}{\sum\limits_{{{M}_{h}},{{\varepsilon }_{m}}}{{}}}S({{M}_{h}},{{\varepsilon }_{m}};\vec{Q})S({{R}_{i}};-\vec{Q}){{{\hat{\rho }}}_{h}}({{M}_{h}},{{\varepsilon }_{m}};\vec{Q}){{{\hat{\rho }}}_{e}}({{R}_{i}};-\vec{Q})- & \nonumber \\
&-\sum\limits_{i}{\sum\limits_{{{M}_{h}},\varepsilon _{m}^{-}}{{}}}S({{M}_{h}},\varepsilon _{m}^{-};\vec{Q})S({{R}_{i}};-\vec{Q}){{{\hat{\rho }}}_{h}}({{M}_{h}},\varepsilon _{m}^{-};\vec{Q}){{{\hat{\rho }}}_{e}}({{R}_{i}};-\vec{Q})\}&
\end{eqnarray}
\end{widetext}
In the concrete variant named as ${{F}_{1}}$, when the electrons are in the state ${{R}_{1}}$, whereas the holes are in the state $\varepsilon _{3}^{-}$ with a given value of ${{M}_{h}}$ the Hamiltonian (72) looks as
\begin{eqnarray}
& {{H}_{Coul}}({{R}_{1}};\varepsilon _{3}^{-})= & \nonumber \\
& =\frac{1}{2}\sum\limits_{{\vec{Q}}}{{}}W(\vec{Q})\{{{(|{{a}_{0}}{{|}^{2}}+|{{b}_{1}}{{|}^{2}}{{A}_{1,1}}(\vec{Q}))}^{2}}\times & \nonumber \\
& \times [{{{\hat{\rho }}}_{e}}({{R}_{1}};\vec{Q}){{{\hat{\rho }}}_{e}}({{R}_{1}};-\vec{Q})-{{{\hat{\rho }}}_{e}}({{R}_{1}};0)]+ & \nonumber \\
& +{{(|d_{0}^{-}{{|}^{2}}+|c_{3}^{-}{{|}^{2}}{{A}_{3,3}}(\vec{Q}))}^{2}}\times  & \nonumber \\
& \times [{{{\hat{\rho }}}_{h}}({{M}_{h}},\varepsilon _{3}^{-};\vec{Q}){{{\hat{\rho }}}_{h}}({{M}_{h}},\varepsilon _{3}^{-};-\vec{Q})-{{{\hat{\rho }}}_{h}}({{M}_{h}},\varepsilon _{3}^{-};0)]- & \nonumber \\
& -2(|{{a}_{0}}{{|}^{2}}+|{{b}_{1}}{{|}^{2}}{{A}_{1,1}}(\vec{Q}))\times & \\
& \times (|d_{0}^{-}{{|}^{2}}+|c_{3}^{-}{{|}^{2}}{{A}_{3,3}}(\vec{Q})){{{\hat{\rho }}}_{e}}({{R}_{1}};\vec{Q}){{{\hat{\rho }}}_{h}}({{M}_{h}},\varepsilon _{3}^{-};-\vec{Q})\} & \nonumber
\end{eqnarray}
In the absence of the RSOC we have ${{a}_{0}}=d_{0}^{-}=1$ and ${{b}_{1}}=c_{3}^{-}=0$. In the variant ${{F}_{1}}=({{R}_{1}},\varepsilon _{3}^{-})$ described by the Hamiltonian (73) the 2D magnetoexciton can be described by the wave function
\begin{equation}
\left| {{\Psi }_{ex}}({{F}_{1}},\vec{K}) \right\rangle =\frac{1}{\sqrt{N}}\sum\limits_{t}{{}}{{e}^{i{{K}_{y}}tl_{0}^{2}}}a_{{{R}_{1}},\frac{{{K}_{x}}}{2}+t}^{\dagger }b_{{{M}_{h}},\varepsilon _{3}^{-},\frac{{{K}_{x}}}{2}-t}^{\dagger }\left| 0 \right\rangle
\end{equation}
where $\left| 0 \right\rangle $ is the vacuum state determined by the equalities
\begin{equation}
{{a}_{\xi ,t}}\left| 0 \right\rangle ={{b}_{\eta ,t}}\left| 0 \right\rangle =0
\end{equation}
In the Ref.[22] there were considered also another seven combinations of the electron and hole states as follows:
\begin{eqnarray}
{{F}_{2}}=({{R}_{2}},\varepsilon _{3}^{-}),{{F}_{3}}=({{R}_{1}},\varepsilon _{0}), \nonumber \\
{{F}_{4}}=({{R}_{2}},\varepsilon _{0}^{{}}),{{F}_{5}}=({{R}_{1}},\varepsilon _{4}^{-} ), \nonumber \\
{{F}_{6}}=({{R}_{2}},\varepsilon _{4}^{-}),{{F}_{7}}=({{R}_{1}},\varepsilon _{1}), \nonumber \\
{{F}_{8}}=({{R}_{2}},\varepsilon _{1})
\end{eqnarray}
In all these cases the exciton creation energies were calculated using the formulas
\begin{eqnarray}
& {{E}_{ex}}({{F}_{n}},\vec{k})={{E}_{cv}}({{F}_{n}})-{{I}_{ex}}({{F}_{n}},\vec{k})& \nonumber  \\
& {{E}_{cv}}({{F}_{n}})-{{E}_{g}}={{E}_{e}}(\xi )+{{E}_{h}}(\eta ),& \nonumber \\
& {{F}_{n}}=(\xi ,\eta )&
\end{eqnarray}
Here ${{E}_{g}}$ is the semiconductor energy gap in the absence of the magnetic field. ${{I}_{ex}}({{F}_{n}},\vec{k})$ is the ionization potential of the magnetoexciton.
The numerical results obtained in the Ref.[22] are represented below in the figures 4, 5, 6 being reproduced from the Fig. 3, 4, 5 of the Ref.[22]. The energies do not depend on the quantum numbers ${{M}_{h}}$ because the Zeeman interaction was not taken into account up till now.
\begin{figure}[h]
\includegraphics[scale=0.55]{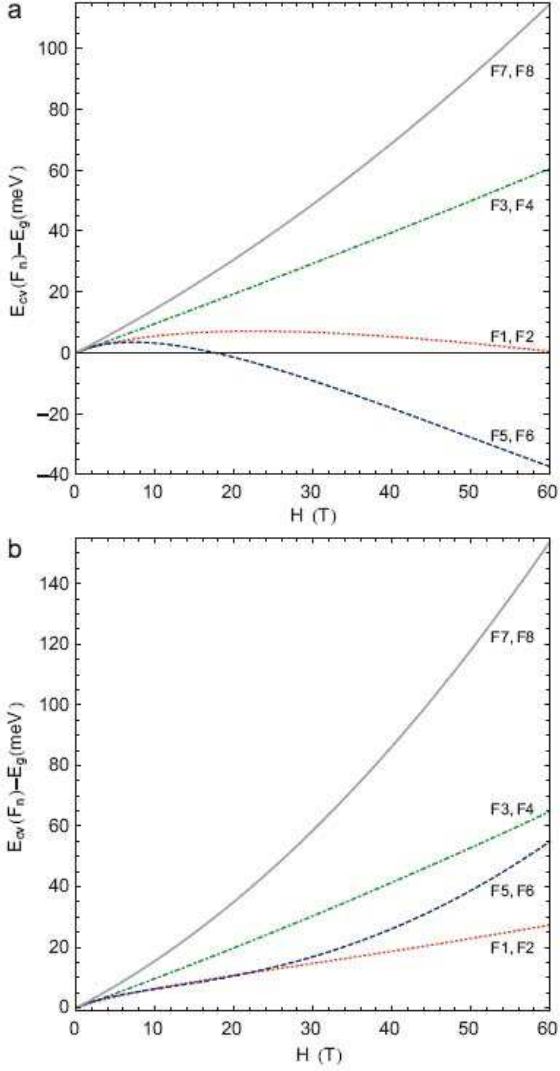}
\caption{The energies ${{E}_{cv}}({{F}_{n}})$ of the band-to-band quantum transitions starting from the LLLs of the heavy holes with the creation of the conduction electrons on the nearly degenerate two LLLs at the parameter ${{E}_{z}}=10\text{kV/cm}$ and two values of the constant C=2.6 (a) and 5.65 (b), reproduced from the Ref.[22].}
\end{figure}

\begin{figure}[h]
\includegraphics[scale=0.55]{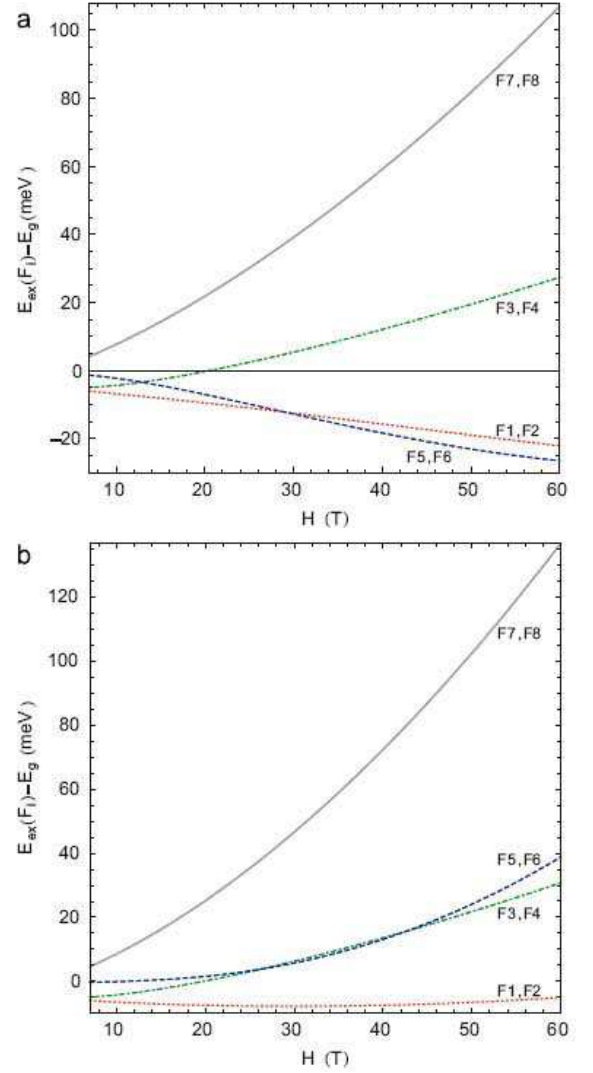}
\caption{The exciton energy levels in dependence on the magnetic field strength $H$ at the parameter ${{E}_{z}}=10\text{kV/cm}$ and two values of the coefficient C: 3.35 (a) and 5.65 (b), reproduced from the Ref.[22].}
\end{figure}

\begin{figure}[h]
\includegraphics[scale=0.55]{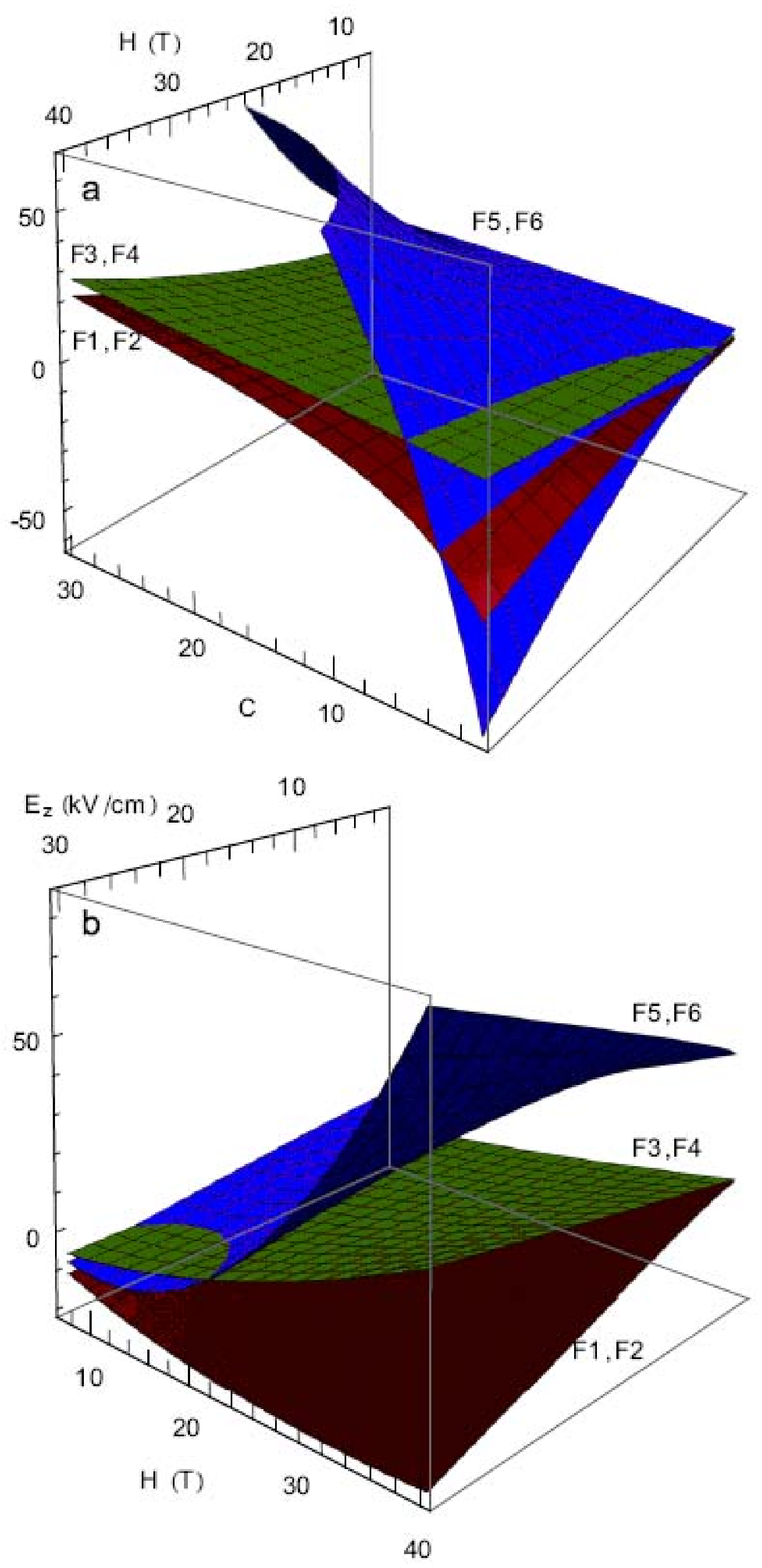}
\caption{Energy spectrum of the lower exciton energy levels with wave vector $\vec{k}=0$ at a given parameter ${{E}_{z}}=10\text{kV/cm}$ and different values of the coefficient C and magnetic field strength $H$ (a), as well as at a given coefficient C=10 and different values of parameters ${{E}_{z}}$ and $H$ (b), reproduced from the Ref.[22].}
\end{figure}
In the papers [24, 25]  on the base of Hamiltonian (72) another aspects of the magnetoexciton physics with special attention devoted to the Rashba spin-orbit coupling were studied. The influence of this interaction on the chemical potential of the Bose-Einstein condensed magnetoexcitons and on the ground-state energy of the metallic-type electron-hole liquid (EHL) in the Hartree-Fock approximation (HFA) was investigated. The magnetoexciton ground-state energy, and the energy of the single-particle elementary excitations were obtained. As is shown in the Fig.7 the chemical potential is a monotonic function versus the value of the filling factor ${{v}^{2}}$ with negative compressibility. It leads to the instability of the Bose-Einstein condensate of magnetoexcitons, when the influence of the excited Landau levels is not taken into account. The energy per one e-h pair in the frame the electron-hole droplets (EHD) was found to be situated on the energy scale lower than the value of the chemical potential of the Bose-Einstein condensed magnetoexcitons with wave vector $\vec{k}=0$ calculated in the HFA. In the Fig.8 the dependence of the energy of the single-particle elementary excitations in condition of BEC of magnetoexcitons on two parameter is presented. One of them is the filling factor ${{v}^{2}}$, but another one is the condensate wave vector $\vec{k}$. With the increasing of the condensate wave vector the energy of the single-particle elementary excitations decreases asymptotically. In the Fig. 9 the ground state energy in condition of BEC of magnetoexcitons is drawn. It has a reverse picture as compared with fig. 8 and is characterized by the increasing with saturation dependence on the condensate wave vector.
\begin{figure}[h]
\includegraphics[scale=0.28]{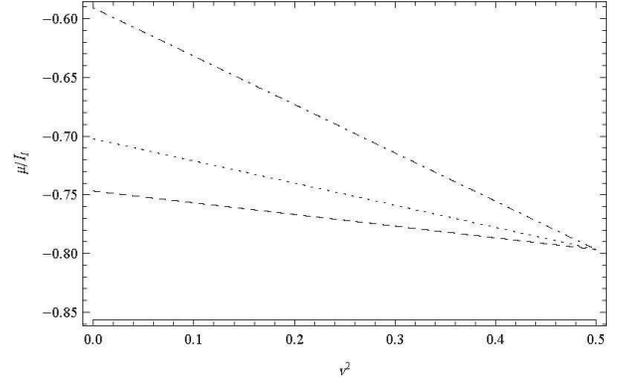}
\caption{Chemical potential in units of exciton binding energy ${{I}_{l}}$ versus filling factor ${{v}^{2}}$. Solid line: energy per e-h pair in EHD phase; dashed line: chemical potential of condensed excitons with $kl=0$; dotted line: the same, but for $kl=0.5$; dash-dotted line: the same, but for$kl=1$, reproduced from the Ref.[25]}
\end{figure}

\begin{figure}[h]
\includegraphics[scale=0.28]{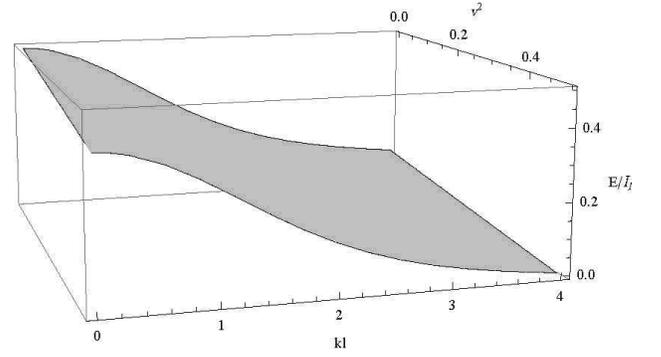}
\caption{The dimensionless energy of the single-particle elementary excitations versus filling factor ${{v}^{2}}$ and in dependence of the condensate wave vector $\vec{k}$, reproduced from the Ref.[25]}
\end{figure}

\begin{figure}[h]
\includegraphics[scale=0.35]{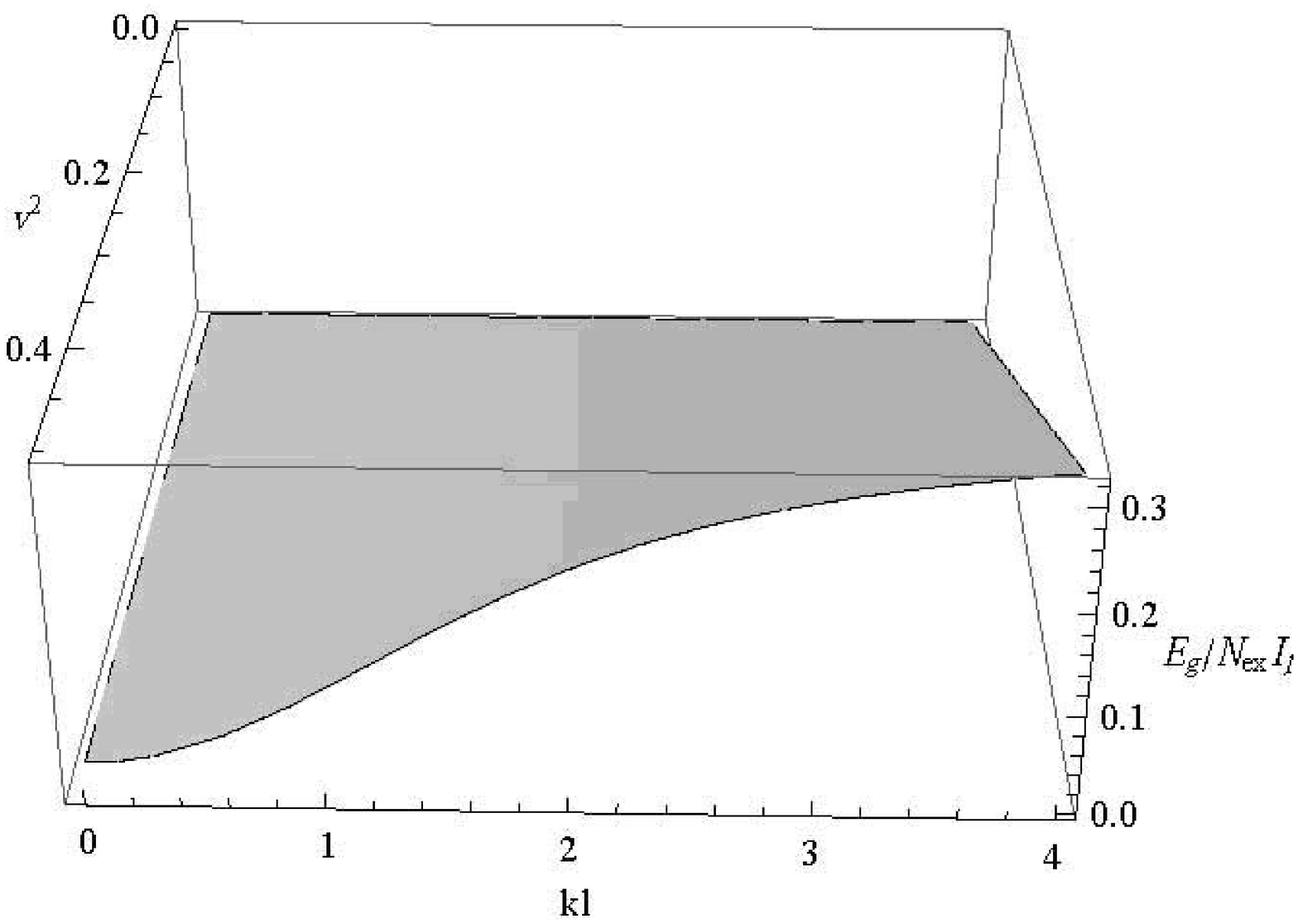}
\caption{The dimensionless ground state energy per one magnetoexciton versus filling factor ${{v}^{2}}$ and in dependence on the condensate wave vector $\vec{k}$, reproduced from the Ref.[25]}
\end{figure}

Some information about the magnetoexciton-polariton formation is needed. This topic was discusses in the Ref.[21, 23] without taking into account the RSOC. It was supposed that the main structure of the 2D magnetoexciton is determined by the strong perpendicular magnetic field which gives rise to the Landau quantization states of the electrons and holes. In the Refs.[19, 22] as well as in the present review article the RSOC effects were considered to be inseparable from the Landau quantization procedure. The Coulomb and electron-radiation interactions take place on this background. In the hierarchy of these four interactions the Coulomb e-h interaction takes the third position if it determines the relative e-h motion in the frame of the magnetoexciton, rather than the electron-radiation interaction. It takes place when the magnetoexciton ionization potential ${{I}_{l}}$ prevails on the Rabi energy $\hbar |\omega_R|$. The Rabi frequency $|{{\omega }_{R}}|$ was determined in Ref.[21] in the case of inter-band dipole-active quantum transitions and equals to
\begin{eqnarray}
& |{{\omega }_{R}}|=\frac{e}{{{m}_{0}}{{l}_{0}}}\sqrt{\frac{1}{{{L}_{c}}\hbar {{\omega }_{{\vec{k}}}}}}|{{P}_{cv}}(0)|{{\delta }_{kr}}({{n}_{e}},{{n}_{h}})\approx  & \nonumber \\
& \approx \sqrt{H}|{{P}_{cv}}(0)|{{\delta }_{kr}}({{n}_{e}},{{n}_{h}}), & \nonumber  \\
& {{\omega }_{{\vec{k}}}}=\frac{c}{{{n}_{c}}}\sqrt{\frac{{{\pi }^{2}}}{L_{c}^{2}}+\vec{k}_{||}^{2}},\vec{k}={{{\vec{a}}}_{3}}{{k}_{z}}+\vec{k}_{||} &
\end{eqnarray}
here ${{l}_{0}}$ is the magnetic length ${{l}_{0}}=\sqrt{\hbar c/eH}$, ${{L}_{c}}$ and ${{n}_{c}}$ are the length and refractive index of the microcavity, ${{P}_{cv}}(0)$ is determined by the formula (30). The Kronekker $\delta $-symbol indicates the selection rule requiring the equality of the numbers ${{n}_{e}}$ and ${{n}_{h}}$ of the electron and hole Landau quantization levels ${{n}_{e}}={{n}_{h}}$. In the Faraday geometry when the light wave vector $\vec{k}$ is oriented along the cavity axis ${{\vec{a}}_{3}}$, the light with a circular polarization ${{\vec{\sigma }}_{k}}$ excites only one magnetoexciton state ${{M}_{h}}$. This effect is known as optical alignment and belongs to the optical orientation phenomena [30].

The matrix element of the band-to-band quantum transition $|{{\vec{P}}_{cv}}(0)|$ may be expressed through the oscillator strength $f_{ex}$ of the optical quantum transition from the ground state of the bulk crystal to the 3D Wannier-Mott exciton state using the formula [28]
\begin{eqnarray}
{{f}_{ex}}=\frac{2}{{{m}_{0}}{{E}_{g}}}|{{{\vec{P}}}_{cv}}(0){{|}^{2}}|{{\Psi }_{ex}}(0){{|}^{2}}{{v}_{0}}, \nonumber \\
 \ |{{\Psi }_{ex}}(0){{|}^{2}}=\frac{1}{\pi a_{ex}^{3}};\ {{v}_{0}}=a_{0}^{3},
\end{eqnarray}
where $E_g$ is the semiconductor energy gap, $v_0$ is the volume of the lattice cell and $\Psi_{ex} (0)$ is the wave function of the relative e-h motion with the Bohr radius $a_{ex}$. If one supposes the parameters of the GaAs-type crystal $E_g \sim 1.5~\mathrm{eV}$, $a_{ex} \sim 10^{-6}~\mathrm{cm}$, $a_0 \sim 2 \times 10^{-8}~\mathrm{cm}$ and $f_{ex} \sim 10^{-6}$, the value $|\vec{P}_{cv} (0)| \approx 2 \times 10^{-20}~\mathrm{g~cm/sec}$ will be found. Together with the parameters of the light $\hbar \omega_k \sim E_g$, resonator length $L_c \approx 4 \times 10^{-5}~\mathrm{cm}$, magnetic field strength $H=6T$ and $l_0 \sim 10^{-6}~\mathrm{cm}$ the Rabi frequency of the order of magnitude $\omega_{R} \sim 10^{12}~\mathrm{sec^{-1}}$ was calculated. At this parameters and ${{\varepsilon }_{0}}=10$ the magnetoexciton ionization frequency is about 1013 1/sec and prevails ${{\omega }_{R}}$ by an order of magnitude.
\section{Conclusions:}
The influence of the RSOC on the properties of the 2D magnetoexcitons was described taking into account the results concerning the Landau quantization of the 2D electrons and holes with nonparabolic dispersion laws, pseudospin components and chirality terms [18, 19, 22]. The main attention was paied to the study of the operators ${{\hat{\rho }}_{e}}(\vec{Q})$ and ${{\hat{\rho }}_{h}}(\vec{Q})$, which together with the magnetoexciton creation and annihilation operators $\hat{\Psi }_{ex}^{\dagger }({{F}_{n}},\vec{k})$ and $\hat{\Psi }_{ex}^{{}}({{F}_{n}},\vec{k})$ form a set of four two-particle integral operators. It was shown that the Hamiltonians of the electron-radiation and Coulomb electron-electron interactions can be expressed in the terms of these four integral two-particles operators. It opens the possibility of more effective investigations of the collective elementary excitation and of the quantum states of the 2D magnetoexcitons and magnetopolaritons [18-25] in the presence of the RSOC.

\end{document}